\documentclass[a4paper,10pt]{article}
\usepackage{graphicx}
\usepackage{amssymb}
\usepackage{amsmath}
\usepackage{dcolumn}
\usepackage{bm}
\usepackage{multirow}
\usepackage{cite}
\usepackage{xcolor}
\usepackage{url}
\usepackage{mathrsfs}
\usepackage[normalem]{ulem}
\usepackage{lscape}
\usepackage{soul}
\usepackage{subcaption} 
\usepackage{dcolumn,ulem,enumitem}
\usepackage{bm}
\usepackage{xcolor}
\usepackage{tikz}
\usepackage{booktabs}
\usepackage{amsfonts}
\usepackage{float}
\usepackage{caption}
\usepackage{subcaption}
\usepackage{mathrsfs}
\usepackage{xcolor}
\usepackage{soul}
\setstcolor{red}
\usepackage{empheq}
\usepackage{hyperref}
\usepackage[utf8]{inputenc}
\usepackage[T1]{fontenc}
\usepackage{lmodern} % Optional: improves font rendering with T1
\usepackage{multicol}

\begin{document}

\huge

\begin{center}
Inferring partial crystalline order in liquids from electrical resistivity
\end{center}

\vspace{0.5cm}

\large

\begin{center}
Nadine Wetta\footnote{nadine.wetta@cea.fr} and Jean-Christophe Pain
\end{center}

\normalsize

\begin{center}
\it CEA, DAM, DIF, F-91297 Arpajon, France\\
\it Universit\'e Paris-Saclay, CEA, Laboratoire Mati\`ere en Conditions Extr\^emes,\\
\it 91680 Bruy\`eres-le-Ch\^atel, France\\
\end{center}

\vspace{0.5cm}

\begin{abstract}
This work investigates how locally persistent crystal-like ordering in liquids influences the Debye-Waller factor. We have developed a theoretical framework based on liquid-phonon theory which introduces a phonon relaxation time, expressed as the ratio of shear viscosity to infinite-frequency shear modulus. These values are obtained using the Yukawa one-component plasma model. Within this framework, we establish expressions for the heat capacity at constant pressure and the Debye–Waller factor for the liquid state. These expressions explicitly introduce additional temperature dependence arising from the finite phonon lifetime. Anharmonicity is accounted for within the quasi-particle approximation. We compare our heat capacity results with values measured by Gathers for aluminum and copper, finding good agreement when assuming partial local crystal-type order. Comparisons with experimental heat capacities serve to validate the approach prior to its application to the study of electrical resistivity, the principal objective of this work. Using liquid-phonon Debye-Waller factors in the methodology developed earlier in [\href{https://journals.aps.org/pre/abstract/10.1103/PhysRevE.102.053209}{Phys. Rev. E {\bf 102}, 053209 (2020)}] for electrical resistivity in dense matter, and comparing with experimental resistivities from Gathers, we elucidate the character of the locally persisting crystal order in liquid aluminum and liquid copper. These results indicate that the electrical resistivity measurements can serve as a valuable probe for determining both the extent and the nature of crystalline order in the liquid state.   
\end{abstract}

\section{Introduction}\label{sec1}

The warm dense matter (WDM) regime represents a unique and complex state of matter that lies at the confluence of condensed matter physics, plasma physics, and the physics of dense liquids. Despite its prevalence in a wide array of physical systems - including the interiors of exoplanets, inertial confinement fusion experiments, and the outer layers of neutron stars - WDM remains both poorly understood and notoriously difficult to characterize. Recent advancements in experimental capabilities now allow for the precise generation and diagnostic probing of WDM conditions, opening unprecedented avenues for exploring its microscopic properties and testing the predictive power of theoretical frameworks. In particular, the advent of X-ray free-electron lasers (XFELs), such as the Linac Coherent Light Source (LCLS), has made it possible to investigate WDM with exceptional temporal and spectral resolution. Through advanced X-ray scattering techniques, key parameters such as electron density, temperature, and ionization states can now be directly measured in compressed materials. The principal theoretical difficulty in modeling WDM lies in its intermediate nature: it is neither amenable to the conventional methods of condensed matter physics nor fully captured by traditional plasma physics. The fundamental challenge arises from the inability to describe its thermodynamic behavior using perturbative approaches around idealized models. Classical plasma approaches break down in the WDM regime due to the significant role of quantum effects, especially those involving electrons. Conversely, extending condensed matter approaches - traditionally limited to low-temperature regimes - into this highly excited domain is both necessary and nontrivial.\\
\indent An efficient alternative strategy for modeling WDM consists in resorting to quantum average-atom (AA) models. These models, which build upon the finite-temperature extension of the semi-classical Thomas–Fermi theory, provide a self-consistent Density Functional Theory (DFT)-based description of atomic structure. Typically formulated within the local density approximation (LDA) to incorporate exchange-correlation effects at finite temperature, these models offer robust convergence - especially at elevated temperatures - and can reliably yield both equations of state and transport properties across a broad spectrum of thermodynamic conditions.\\
\indent A few years ago, we developed a methodology that enables a unified and physically consistent computation of the electrical conductivity of dense matter across a wide range of regimes - from the solid state to hot plasma - using a single theoretical framework \cite{Wetta2020}. This approach is grounded in the Ziman theory, where we systematically exclude elastic scattering contributions to the structure factors of solids and liquids. For the solid phase, the structure factor is evaluated through a multiphonon expansion. The elastic scattering component corresponds to the zero-phonon term, manifesting as Bragg peaks that are thermally suppressed via Debye-Waller attenuation. Bragg scattering of electrons, which gives rise to the formation of energy band structures through the emergence of Bloch states, does not contribute to the electron-ion collision integral in the kinetic equation. This is because such coherent, elastic scattering processes preserve crystal momentum and do not lead to net momentum transfer or energy dissipation relevant to electrical resistivity. Such an idea was first evoked by astrophysicists \cite{Flowers1976} and originally applied to solids by Rosenfeld and Stott \cite{Rosenfeld1990}. Their approach involves subtracting the contribution of a perfectly rigid lattice from the total structure factor used in Ziman's formula, thereby isolating the inelastic components relevant for transport properties. Baiko \emph{et al.} \cite{Baiko1998} adapted it to astrophysical contexts, demonstrating its broader applicability beyond crystalline solids. Indeed, numerical simulations conducted by various authors have revealed the emergence of incipient long-range order in systems where the Coulomb coupling parameter is $\Gamma\gg 1$. For instance, Schmidt \emph{et al.} \cite{Schmidt1997} reported the presence of a shear mode at $\Gamma > 100$ in their molecular dynamics simulations, which coexist with the well-known longitudinal ion-acoustic (plasmon) mode. The analysis of Baiko \emph{et al.} confirms that the spectral features of these collective modes can be effectively interpreted as an angular average of the phonon dispersion relations characteristic of a crystalline lattice. Although this long-range order is inherently transient and eventually decays, it can persist over timescales comparable to the electron scattering time. As a consequence, a temporary electron band structure can form, analogous to that in a solid, during which elastic electron scattering does not contribute to electrical resistance. This interpretation aligns with the perspective advanced by Edwards \cite{Edwards1962}, who argued that transport properties should be governed by the local structural order experienced by an electron along its mean free path, rather than by the global disorder of the system.\\
\indent Relevant models of the Debye-Waller (DW) factors are needed to correctly describe the thermal attenuation of the elastic scattering part of the total structure factor. Potekhin \emph{et al.} \cite{Potekhin1999} derived the formula for the plasma DW based on the assumption of a Coulomb potential screened by the static polarization of ideal, relativistic, strongly degenerate electrons (the authors were mostly interested in the interior of white dwarfs and neutron stars). They proposed an interpolation formula between the plasma DW and that of a solid. We used it in Ref.~\cite{Wetta2020} for our calculations, including those in the liquid state, which typically falls within the interpolation area. \\
\indent The present work focuses on deriving Debye-Waller factors specific to liquids, thus avoiding the need for interpolation between two limiting forms. Our approach is based on the liquid-phonon theory developed by Bolmatov \emph{et al.} \cite{Bolmatov2011,Bolmatov2012,Bolmatov2013,Bolmatov_Zhernenkov2015,Bolmatov_Zavyalov2015,Bolmatov2022}. We resort to the Yukawa One-Component Plasma (YOCP) model \cite{Hubbard1971,Baus1980} to obtain quantities, such as the shear viscosity, the instantaneous shear modulus, the Gr\"uneisen parameter or the thermal expansion coefficient, required by the liquid-phonon theory. Anharmonic effects are introduced within the quasi-particle (QP) approximation \cite{Allen2015,Allen2020}.\\
\indent The liquid-phonon theory used in this study is presented along with the QP approach for anharmonicity in Sec.~\ref{sec2}, while the YOCP formulas used for the quantities needed by this theory are given in Sec.~\ref{sec3}. Before applying this approach to the modeling of the DW factors in liquids, we test it in Sec.~\ref{sec4} by comparing with available heat capacities at constant pressure $C_P(T)$ and volume expansion curves $V(T)$. Comparison with experimental shear viscosities and shear moduli provides insight into the relevance of the YOCP quantities. In Sec.~\ref{sec5}, we use liquid-phonon theory to calculate DW factors in liquids. We then apply them in our calculations of the electrical resistivity of liquid aluminum and liquid copper, following the formalism we developed in Ref.~\cite{Wetta2020}, that we recall in Sec.\ref{subsec51}. Our results show that the liquid-phonon Debye-Waller factors provide a more detailed means of inferring information about the local atomic environment in these liquids from experimental data than the interpolation formula of Potekhin \emph{et al.}. 

\section{Liquid-phonon theory}\label{sec2}

\subsection{Frenkel's vision of the liquid state}\label{subsec21}

The understanding of the thermodynamics of the liquids is a very hard task, at least much more complicated than for gases and solids. The main difficulty stems from the fact that, in a liquid, interactions are strong and also depend on the type of liquid itself. Previously, liquid potential energy has been calculated from correlation functions and interatomic interactions. Several approaches were developed for this purpose \cite{Landau2013,March1990,Lebowitz1963,Curtin1985,Rosenfeld1986,Bolmatov2009}. Concretely, however, these theories only successfully apply to the simple liquids, which are characterized by short-range order and two-body correlations, and become rapidly intractable in more complex ones. A phonon solid-state approach is, at first sight, equally impossible. The Debye model is very effective in explaining the solid's heat capacity, no matter how interactions and correlations are, but it relies on the assumption of small atomic displacements, allowing internal energy to be developed in terms of small squared atomic displacements. In a liquid, the latter are much stronger, and such a method becomes difficult to justify. Finally, approaching the liquid state from the gas phase is not more relevant, due to the fact that interatomic interactions are much stronger than in gases, and should not be considered as perturbations.

An alternative approach was put forward as early as the mid-twentieth century by Frenkel \cite{Frenkel1947}, but was only seriously considered later, due to the lack of experimental evidence and of mathematical rigorous justifications. Frenkel's microscopic picture of liquids assumes that local order persists for a time $t\leq\tau$ in liquids, after which it relaxes. For $t\leq\tau$, the liquid can then be considered able to support shear-waves. Beyond $t=\tau$, atoms diffuse, favored by rearrangements, at the expense of the ability to respond to shear waves. Shear waves can therefore exist in liquids beyond a cutoff frequency $\omega_F$ given by the inverse of the relaxation time $\tau$. According to the Maxwell model of the damped harmonic oscillator, the relaxation time is the ratio $\tau=\eta/G_\infty$ of the viscosity $\eta$ to the infinite-frequency shear modulus $G_\infty$. 

From the intuitive standpoint that, if phonon states in liquids only depend on $\tau$, the thermodynamic properties of liquids should also only depend on it, Bolmatov \emph{et al.} followed Frenkel's idea and developed a phonon theory applicable to liquid thermodynamics, in both classical and quantum regimes \cite{Bolmatov2011,Bolmatov2012,Bolmatov2013,Bolmatov_Zavyalov2015,Bolmatov_Zhernenkov2015,Trachenko2008}. They obtained good agreement between their calculations and experimental heat capacities for 21 liquids, including complex ones, like molecular and hydrogen-bonded network liquids, in wide ranges of temperature and pressure \cite{Bolmatov2012}. Thus the work of Bolmatov \emph{et al.} tends to indicate that understanding the thermodynamical properties of liquids may be easier than previously expected, despite the complexity of their inherent interactions and correlations.

\subsection{Bolmatov \emph{et al.}'s derivation of the energy of the liquid}\label{subsec22}

In the following, we present the main lines of Bolmatov \emph{et al.}'s derivation of the energy \cite{Bolmatov2012}. The authors started by expressing the energy of the liquid as the sum of vibration and diffusion terms as
\begin{equation*}
    E=E_\mathrm{vib}+E_\mathrm{dif}.
\end{equation*}
The vibration energy $E_{vib}$ is itself the sum of the contributions of longitudinal and shear waves, respectively indicated by subscripts $||$ and $\perp$. Making the distinction between kinetic $\mathcal{K}$ and potential $\mathcal{P}$ energies, and remembering that only shear waves with frequencies $\omega\geq\omega_F$ persist
\begin{equation*}
    E_{vib}=\mathcal{K}_{||}+\mathcal{P}_{||}+\mathcal{K}_\perp(\omega\geq\omega_F)+\mathcal{P}_\perp(\omega\geq\omega_F).
\end{equation*}
Similarly, the diffusion energy is the sum of kinetic and potential terms, \emph{i.e.} $E_{dif}=\mathcal{K}_{dif}+\mathcal{P}_{dif}$, resulting in the following expression for the liquid's energy:
\begin{equation*}
    E=\mathcal{K}_{||}+\mathcal{P}_{||}+\mathcal{K}_\perp(\omega\geq\omega_F)+\mathcal{P}_\perp(\omega\geq\omega_F)+\mathcal{K}_\mathrm{dif}+\mathcal{P}_\mathrm{dif}.
\end{equation*}
In Frenkel's picture of liquids, the vibrational shear motion with $\omega<\omega_F$ is replaced by the diffusive motion during the time interval $\tau$. Therefore, $\mathcal{P}_\mathrm{dif}$ is comparable to $\mathcal{P}_\perp(\omega<\omega_F)$. Since $\mathcal{P}_\perp(\omega<\omega_F)\ll \mathcal{P}_\perp(\omega\geq\omega_F)$, $\mathcal{P}_\mathrm{dif}$ can be neglected. After summing all the kinetic contributions in $\mathcal{K}$, one obtains:
\begin{equation*}
    E=\mathcal{K}+\mathcal{P}_{||}+\mathcal{P}_\perp(\omega\geq\omega_F).
\end{equation*}
According to the equipartition theorem: $\mathcal{K}=\frac{E_{||}+E_\perp}{2}$, $\mathcal{P}_{||}=\frac{E_{||}}{2}$, and $\mathcal{P}_\perp(\omega\geq\omega_F)=\frac{E_\perp(\omega\geq\omega_F)}{2}$. Then:
\begin{align}\label{E_liquid_phonon}
    E&=E_{||}+\dfrac{E_\perp}{2}+\dfrac{E_\perp(\omega\geq\omega_F)}{2}\nonumber\\
     &=E_{||}+\dfrac{E_\perp(\omega<\omega_F)+E_\perp(\omega\geq\omega_F)}{2}+\dfrac{E_\perp(\omega\geq\omega_F)}{2}\nonumber\\
     &=E_{||}+E_\perp(\omega\geq\omega_F)+\dfrac{E_\perp(\omega<\omega_F)}{2}.
\end{align}
The first two terms of the last equality correspond to the elastic contributions of the longitudinal phonon modes and the transverse frequencies $\omega\geq\omega_F$ that still persist in the liquid state. The third term reflects the “conversion” to viscous energy of the elastic energy due to shear waves with frequencies $\omega\leq\omega_F$ that disappear in the liquid state.

Within the Debye model, we have
\begin{equation*}
E_{||}=\int_0^{\omega_D} E(\omega) g_{||}(\omega) d\omega,
\end{equation*}
with
\begin{equation*}\label{E_omega}
    E(\omega)=\dfrac{\hbar\omega}{2}+\dfrac{\hbar\omega}{\exp(\beta\hbar\omega)-1}, 
\end{equation*}
where $\omega_D$ denotes the Debye frequency. $g_{||}(\omega)$ is the longitudinal phonon density of states, given, in the Debye model (for $N$ atoms), by
\begin{equation*}
    g_{||}(\omega)=3N \dfrac{\omega^2}{\omega_D^3}.
\end{equation*}
In the same way, one has
\begin{equation*}
    E_\perp(\omega\geq\omega_F)= \int_{\omega_F}^{\omega_D} E(\omega) g_\perp(\omega)\,\mathrm{d}\omega, 
\end{equation*}
and
\begin{equation*}
E_\perp(\omega<\omega_F)= \int_{0}^{\omega_F} E(\omega) g_\perp(\omega)\,\mathrm{d}\omega, 
\end{equation*}
where $g_\perp(\omega)$ denotes the transverse phonon density of states, given by
\begin{equation*}
    g_\perp(\omega)=6N \dfrac{\omega^2}{\omega_D^3}.
\end{equation*}
Introducing these expressions into Eq.~(\ref{E_liquid_phonon}) gives
\begin{equation*}\label{E_LPT}
    E=3N \int_0^{\omega_D} E(\omega) \dfrac{3\omega^2}{\omega_D^3} d\omega- N \int_0^{\omega_F} E(\omega) \dfrac{3\omega^2}{\omega_D^3}\,\mathrm{d}\omega. 
\end{equation*}
Bringing in the ``Frenkel temperature'' $\theta_F=\hbar\omega_F/k_B$, the Debye one $\theta_D=\hbar\omega_D/k_B$ and the 3-dimensional Debye function defined by
\begin{equation*}\label{D3}
    D_3(x)=\dfrac{3}{x^3}\int_0^x\dfrac{t^3}{e^t-1}\,\mathrm{d}t, 
\end{equation*}
the energy reads, within the Debye theory of the harmonic lattice
\begin{equation*}
    E=E_0+Nk_BT\,\left[3D_3\left(\dfrac{\theta_D}{T} \right)-\left(\dfrac{\theta_F}{\theta_D}\right)^3 D_3\left(\dfrac{\theta_F}{T} \right) \right],
\end{equation*}
where $E_0$ is the so-called zero-point vibration energy
\begin{equation*}
    E_0=\dfrac{3}{8} Nk_B\theta_D\,\left[3-\left(\dfrac{\theta_F}{\theta_D}\right)^4\right]. 
\end{equation*}
The Debye function $D_3(x)$ has an analytic expression in terms of polylogarithm functions $\mathrm{Li}_n(x)$ \cite{Lewin1981,Gonzales2022}:
\begin{equation*}
    D_3(x)=\dfrac{18}{x^3} \zeta(4)-\dfrac{18}{x^3}\mathrm{Li}_4(e^{-x})-\dfrac{18}{x^2}\mathrm{Li}_3(e^{-x})\nonumber\\
    -\dfrac{9}{x}\mathrm{Li}_2(e^{-x})-3\mathrm{Li}_1(e^{-x}),
\end{equation*}
where $\zeta(4)=\pi^4/90$ and $\mathrm{Li}_1(x)=\ln(1-x)$.\\

\subsection{Application to the heat capacities at the quasi-harmonic approximation}\label{subsec23}

\subsubsection{Heat capacity at constant volume $C_V(T)$}

The isochoric heat capacity is the derivative of the energy given in Eq.~(\ref{E_LPT}) with respect to the temperature. 
In the quasi-harmonic (QH) approximation, it reads
\begin{equation}\label{CVh}
    C_V(T)= 4Nk_B\,\left[3D_3\left(\dfrac{\theta_D}{T} \right)-\left(\dfrac{\theta_F}{\theta_D}\right)^3 D_3\left(\dfrac{\theta_F}{T} \right) \right]\nonumber\\
    -3Nk_B \left[3\dfrac{\theta_D/T}{\exp(\theta_D/T)-1}-\left(\dfrac{\theta_F}{\theta_D}\right)^3 \dfrac{\theta_F/T}{\exp(\theta_F/T)-1} \right].
\end{equation}
In the classical limit $\theta_F\leq \theta_D\ll T$, the argument of the Debye function tends to zero, then $D_3(x)\approx 1-\frac{3}{8}x$, and we have
\begin{equation*}
    C_V\rightarrow Nk_B \left[3-\left(\dfrac{\theta_F}{\theta_D}\right)^3\right].
\end{equation*}
For solids, $\theta_F=0$ and the classical Dulong and Petit law $C_V=3Nk_B$ is recovered. For a perfect liquid (in which there is no order other than ionic correlations) $\theta_F=\theta_D$,  and $C_V(T)$ approaches the value $C_V=2Nk_B$ expected for these liquids.

It should be noted that, in order to arrive at Eq.~(\ref{CVh}), it is necessary to assume that the temperature dependence of $\omega_F$ can be neglected. In fact, this dependence introduces an additional contribution in the right-hand side of Eq.~(\ref{CVh}), that reads, at the classical limit
\begin{equation*}
\Delta C_V=-3Nk_B\,\left(\dfrac{\theta_F}{\theta_D}\right)^3\,\left.\dfrac{\partial\ln\omega_F}{\partial\ln T} \right|_V.
\end{equation*}
The infinite-frequency shear modulus $G_\infty$ is a purely elastic quantity that can be assumed to vary in a similar way to vibration frequencies. As a consequence, at the QH approximation: $\left.\frac{\partial\ln\omega_F}{\partial\ln T} \right|_V\approx -\left.\frac{\partial\ln\eta}{\partial\ln T} \right|_V$. The assumption $\Delta C_V\approx 0$ is based on the observation that viscosity follows a strongly decreasing slope (exponential in $1/T$) in the vicinity of melting, then compensated for by the factor $\left(\theta_F/\theta_D\right)^3\approx 0$. Concomitantly, above $T_m$ where this factor rises, the approximation is justified because the experiments show that the viscosity rapidly stabilizes at a fairly constant value (see for instance Refs.~\cite{Assael2006,Assael2010,Assael2012,Assael2018}).

\subsubsection{Heat capacity at constant pressure $C_P(T)$}

The isobaric heat capacity is the temperature derivative of the enthalpy at constant pressure:
\begin{equation*}
    C_P=\dfrac{\partial H}{\partial T}\Bigg|_P=T\dfrac{\partial S}{\partial T}\Bigg|_P.
\end{equation*}
Starting from the following relation
\begin{equation*}
    dS=\left.\dfrac{\partial S}{\partial T}\right|_V dT+ \left.\dfrac{\partial S}{\partial V}\right|_T dV, 
\end{equation*}
one obtains
\begin{align*}
    C_P(T)=T \left.\dfrac{\partial S}{\partial T}\right|_P &= T \left[ \left.\dfrac{\partial S}{\partial T}\right|_V + \left.\dfrac{\partial S}{\partial V}\right|_T\,\left.\dfrac{\partial V}{\partial T}\right|_P \right]\nonumber\\
    &= \left[C_V(T) + \left.\dfrac{\partial P}{\partial T}\right|_V\,\alpha_V T\ V\right],
\end{align*}
where we have introduced the volume expansion coefficient, defined by
\begin{equation*}\label{alphaV_def}
    \alpha_V=\dfrac{1}{V}\left.\dfrac{\partial V}{\partial T}\right|_P.
\end{equation*}
In the QH approximation, the volume thermal expansion parameter $\alpha_V$ is related to the isochoric heat capacity $C_V(T)$ and the isothermal bulk modulus $B_T=-V\left.\frac{\partial P}{\partial V}\right|_T$ by the Gr\"uneisen equation
\begin{equation*}\label{alpha_V}
    \alpha_V=\dfrac{\gamma_G C_V}{B_TV},
\end{equation*}
where $\gamma_G$ is the macroscopic Gr\"uneisen parameter defined by
\begin{equation}\label{gammaG_def}
    \gamma_G=-\left.\dfrac{\partial\ln\omega}{\partial\ln V}\right|_T. 
\end{equation}
Finally, the QH isobaric heat capacity reads
\begin{equation*}\label{Cp}
    C_P^\mathrm{QH}(T)=(1+\gamma_G\,\alpha_V\,T)\,C_V(T),
\end{equation*}
with $C_V(T)$ given in Eq.~(\ref{CVh}). 

\subsection{Beyond the quasi-harmonic approximation}\label{subsec24}

\subsubsection{Allen's quasi-particle approach}

The QH approximation has proven to be a powerful tool for predicting the thermodynamic properties of materials under conditions close to normal. However, its limitations become evident at elevated temperatures, especially under low-pressure conditions. Notably, it fails to describe dynamically stabilized phases and often yields inaccurate results because of its treatment of vibrational frequencies. In the QH approximation, temperature effects are introduced only indirectly through volume expansion (phonon frequencies vary with temperature dependent volume $V(T)$), an approach valid only to first order. As a result, systems featuring soft phonon modes or light atomic masses are particularly vulnerable to these inaccuracies, with the breakdown temperature of QH depending on both the material and the pressure. To address these deficiencies, it is necessary to incorporate anharmonic effects by considering higher-order terms in the vibrational Hamiltonian. This introduces additional complexity, as no exact statistical mechanics solution exists beyond the harmonic approximation. One widely adopted strategy is the use of molecular dynamics, either classical or ab initio, though such simulations are computationally demanding. Machine learning approaches can help mitigate this cost, but the computational burden remains significant \cite{Blancas2024}.

Vibrational spectroscopy studies of sufficiently pure crystals often reveal sharp Lorentzian peaks, which can be assigned to specific wave vectors $Q$. These peaks typically correspond directly to harmonic normal modes and are interpreted as quasi-particles (QPs). The central frequency $\omega_Q$, representing the energy of a QP, is temperature dependent. Comparisons between theoretical predictions and experimental results indicate that the quasi-harmonic (QH) energy $\hbar\omega_Q(V)$, evaluated at thermally expanded volume $V(T)$, fails to accurately reproduce the observed temperature-induced frequency shifts at elevated temperatures. Allen's quasi-particle theory is grounded in two central ideas \cite{Allen2015,Allen2020}. First, low-energy excitations behave like non-interacting particles, each characterized by a QP energy $\hbar\omega_Q(V, T)$ and an associated mode occupancy $\langle \hat{n}_Q \rangle$. Second, the dynamics of these low-energy states can be described by the time- and space-dependent evolution of their occupancies. Nevertheless, this model has its limitations. QP theory can break down when the spectral peak deviates significantly from a Lorentzian shape, making $\omega_Q$ ill-defined, and when mode occupancies become strongly correlated (nonzero value of $\langle \hat{n}_Q \hat{n}_{Q'} \rangle - \langle \hat{n}_Q \rangle \langle \hat{n}_{Q'} \rangle$), violating the non-interacting assumption.

In the context of Allen's theory, $n_Q$ denotes the QP equilibrium Bose-Einstein occupancy, expressed as $n_Q = \left[ \exp(\omega_Q(V,T)/k_B T) - 1 \right]^{-1}$, while $n_{Q,H}$ refers to its harmonic (QH) counterpart, which uses the harmonic frequency $\omega_{Q,H}$. Entropy plays a pivotal role in QP theory. Specifically, it can be interpreted as $S = k_B \log \Omega$, where $\Omega$ is the number of microstates that can distribute a total excitation energy of $\langle n \rangle N \hbar \omega$ across $N$ oscillators. The entropy 
\begin{equation*}
    S = k_B \sum_Q \left[ \left( \langle \hat{n}_Q \rangle + 1 \right) \ln \left( \langle \hat{n}_Q \rangle + 1 \right) - \langle \hat{n}_Q \rangle \ln \langle \hat{n}_Q \rangle \right]
\end{equation*}
is minimized at fixed energy when $\langle\hat{n}_Q\rangle=n_Q$. When harmonic frequencies $\omega_{Q,H}$ are used in $n_Q$, this yields the harmonic entropy $S_H$. Substituting temperature-dependent QP frequencies $\omega_Q(V,T)$ into the expression of the harmonic entropy produces an improved estimate of thermodynamic entropy, denoted $S_\mathrm{QP}$. But inserting QP frequencies into the harmonic free energy $F_H$ results in a free energy $F$ that violates the thermodynamic relation $F = U - T S$, where $S = S_\mathrm{QP}$. Instead, a thermodynamically consistent QP free energy $F_\mathrm{QP}$, which satisfies this relationship, must be explicitly constructed.
 
In harmonic theory, the internal energy is given by $U_H = \sum_Q \omega_{Q,H} (n_{Q,H} + 1/2)$. When anharmonic corrections shift the frequencies to $\omega_Q(V,T)$, the internal energy also requires correction. The leading-order correction arising from second-order perturbation theory involving third- and fourth-order anharmonic force constants reads \cite{Allen2015}
\begin{equation*}\label{UQP}
    U_\mathrm{QP} = \sum_Q \hbar \omega_Q(V,T) \left( n_Q + \frac{1}{2} \right)\nonumber\\
    - \frac{1}{2} \sum_Q \hbar \left[ \omega_Q(V,T) - \omega_{Q,H} \right] \left( n_Q + \frac{1}{2} \right).
\end{equation*}
To this order of the perturbation, the anharmonic frequency shift is given by \cite{Wallace1972,Cowley1963}
\begin{equation*}
\omega_Q(V,T) - \omega_{Q,H} \equiv \Delta\omega^{(2)}_Q = \frac{1}{N} \sum_{Q'} \left( \frac{\partial \omega_Q}{\partial n_{Q'}} \right)_0 \left( n_{Q'} + \frac{1}{2} \right).
\end{equation*}

This correction compensates for the double-counting of interactions and involves a term $\partial \omega_Q / \partial n_{Q'}$, a temperature-independent anharmonic interaction parameter. 
The resulting QP internal energy $U_\mathrm{QP}$  leads to a consistent and accurate free energy:
\begin{equation*}
F_\mathrm{QP} = U_\mathrm{QP} - T S_\mathrm{QP},
\end{equation*}
the thermal part of which is given by
\begin{equation*}
\Delta F_\mathrm{QP}=-\dfrac{\hbar}{2} \sum_Q \Delta\omega_Q^{(2)}  \left(n_Q+\dfrac{1}{2}\right).
\end{equation*}
Another important point is that the anharmonic frequency shift does not vanish at zero temperature. Even at $T = 0$, the QP frequencies differ from harmonic ones due to zero-point motion (since $n_Q + 1/2 \rightarrow 1/2$). This highlights that QP corrections are essential for capturing the true vibrational behavior of a system, even in the quantum limit.

\subsubsection{The QP expressions of $C_V(T)$ and $C_P(T)$}

In the previous section, we followed Allen's notations. In the context of the Debye model used in our work, we will replace the $Q$ indexing of phonon modes used by Allen with frequencies. The sums over discrete reciprocal vectors will become integrals over continuous frequencies (from 0 to $\omega_D$). The harmonic frequencies will be denoted $\omega$, the QH ones $\omega(V)$ and $\omega(V,T)$ will be used for the QPs. With this, the contribution of a given vibration mode to the QP heat capacity at a constant quantity $X$ (which can be the volume $V$ or the pressure $P$), established by Allen \cite{Allen2015}, reads 
\begin{equation*}
    C_X[\omega(V,T)]=\hbar\omega(V,T)\,\left.\dfrac{\partial n}{\partial T}\right|_H \left[1-\dfrac{T}{\omega(V,T)}\left.\dfrac{\partial \omega(V,T)}{\partial T}\right|_X\right].
\end{equation*}
The subscript $H$ on the derivative $\partial n/\partial T$ means that the Bose-Einstein function $n=[\exp(\beta\hbar\omega)-1]^{-1}$ is derived with respect of the explicit temperature in $\beta=1/(k_BT)$, but not the implicit one of the QP frequency. Thus:
\begin{equation*}
    \left.\dfrac{\partial n}{\partial T}\right|_H=\dfrac{\hbar\omega(V,T)}{k_BT^2} n[\omega(V,T)]\left(n[\omega(V,T)]+1\right). 
\end{equation*}
Assuming that $k_BT\gg\hbar\omega(V,T)$, which is relevant for most liquids, we get
\begin{align*}
    \hbar\omega(V,T)\left.\dfrac{\partial n}{\partial T}\right|_H \approx k_B(\beta\hbar\omega)^2 n(\omega)\left[n(\omega)+1\right],
\end{align*}
where the quantity in the right-hand side is the contribution $C_V(\omega)$ of the phonon mode of frequency $\omega$ to the isochoric harmonic heat capacity. Consequently:
\begin{equation*}
    C_V[\omega(V,T)]=C_V(\omega) \left[1-\dfrac{T}{\omega(V,T)}\left.\dfrac{\partial \omega(V,T)}{\partial T}\right|_V\right],
\end{equation*}
and
\begin{equation*}
    C_P[\omega(V,T)]=C_V(\omega) \left[1-\dfrac{T}{\omega(V,T)}\left.\dfrac{\partial \omega(V,T)}{\partial T}\right|_P\right].
\end{equation*}

Anharmonicity induces a thermal shift $\Delta B_T(V,T)$ in the isothermal bulk modulus $B_T(V,T)=B_T(V[T])+\Delta B_T(V,T)$, where, at the QP level corresponding to the lowest anharmonic terms: 
\begin{equation*}
    \Delta B_T(V,T)\approx T\left.\dfrac{\partial B_T}{\partial T} \right|_{V(T)}\approx T\left.\dfrac{\partial B_{ph}}{\partial T} \right|_{V(T)}, 
\end{equation*}
$B_T(V,T)$ denoting the QP bulk modulus, and $B_T(V[T])\equiv B_T(V,0)$ the QH one. At high temperature we have
\begin{equation*}
    B_{ph}(V,T)=\dfrac{\gamma_G (N+N_1) k_BT}{V},
\end{equation*}
where, following Bolmatov and Trachenko's notations \cite{Bolmatov2011}, $N$ and $N_1$ represent respectively the numbers of longitudinal phonon modes and of shear modes remaining in the liquid state. Introducing the QP thermal expansion parameter $\alpha_V\equiv\alpha_V^\mathrm{QP}$, one has
\begin{equation*}
    V=V(T)\left(1+\alpha_VT\right),
\end{equation*}
and
\begin{align*}
    B_{ph}(V,T)&\approx \dfrac{\gamma_G(N+N_1)k_BT}{V(T)}\left(1-\alpha_VT\right)\\
               &\approx B_{ph}(V[T])-\alpha_VT\,B_{ph}(V[T]),
\end{align*}
from which one gets
\begin{equation*}
    \dfrac{1}{B_{ph}}\left.\dfrac{\partial B_{ph}}{\partial T}\right|_{V}=-\alpha_V.
\end{equation*}
Furthermore, the contribution of each frequency $\omega$ is
\begin{equation*}
    B_{ph}(\omega) =-V\left.\dfrac{\partial P_{ph}(\omega)}{\partial V}\right|_T = -V \left.\dfrac{\partial P_{ph}(\omega)}{\partial \omega}\right|_T \left.\dfrac{\partial \omega}{\partial V}\right|_T,
\end{equation*}
or, using the definition of the Gr\"uneisen constant $\gamma_G$
\begin{equation*}
    B_{ph}(\omega)=\gamma_G~\omega\,\left.\dfrac{\partial P_{ph}(\omega)}{\partial\omega}\right|_T, 
\end{equation*}
which relates the bulk vibration modulus to the vibrational spectra of the material \cite{Liu1992}. In the limit $\beta\hbar\omega\ll1$ one has
\begin{equation*}
    P_{ph}(\omega)\approx \dfrac{\gamma_G}{V}\left(k_BT+\dfrac{(\hbar\omega)^2}{12}\right),
\end{equation*}
as well as 
\begin{equation*}
    B_{ph}(\omega)=\gamma_G\omega\dfrac{\partial P(\omega)}{\partial\omega}\propto \omega^2,
\end{equation*}
\begin{equation*}
    \dfrac{1}{B_{ph}} \left.\dfrac{\partial B_{ph}}{\partial T}\right|_V = \dfrac{2}{\omega} \left.\dfrac{\partial\omega}{\partial T} \right|_V, 
\end{equation*}
and finally
\begin{equation*}
    \left.\dfrac{\partial\omega(V,T)}{\partial T}\right|_V=-\dfrac{\alpha_V}{2} \omega. 
\end{equation*}
This is identical to the expression obtained by Bolmatov and Trachenko through another method, based on the assumption that
$B\propto -T$ \cite{Bolmatov2011}.

The isochoric heat capacity in the QP approximation, after integration over harmonic frequencies, reads
\begin{equation*}
    C^\mathrm{QP}_V(T)=C_V(T)\,\left(1+\dfrac{\alpha_V}{2}T\right).
\end{equation*}

The QP isobaric heat capacity requires the derivation of the frequency with respect to $T$ at constant pressure $P$
\begin{align*}
    \left.\dfrac{\partial \omega(V,T)}{\partial T}\right|_P =&\left.\dfrac{\partial \omega(V,T)}{\partial T}\right|_V + \left.\dfrac{\partial \omega(V,T)}{\partial V}\right|_T  \left.\dfrac{\partial V}{\partial T}\right|_P\\
    =&- \left(\dfrac{1}{2}+\gamma_G \right) \alpha_V\ \omega(V,T),
\end{align*}
giving the following expression for the isobaric heat capacity in the QP approximation
\begin{equation*}
    C_P^\mathrm{QP}(T)= C_V(T)\,\left[1+\alpha_V T\left(\dfrac{1}{2}+\gamma_G\right)\right].
\end{equation*}

\subsubsection{The QP thermal expansion parameter}

The QP approximation also introduces an additional term into the expression of the thermal expansion parameter \cite{Allen2020}:
\begin{equation*}
    \alpha_V\equiv\alpha_V^\mathrm{QP}(T)=\alpha_V^\mathrm{QH}(T)+\Delta\alpha_V^\mathrm{QP}(T),   
\end{equation*}
where $\alpha_V^\mathrm{QH}$ is the QH volume thermal expansion parameter. For the Debye solid, it is given by the Gr\"uneisen equation that reads
\begin{equation*}
    \alpha_V^\mathrm{QH}(T)=\dfrac{\gamma_G C_V}{B_T V},
\end{equation*}
where $B_T=V\left.\frac{\partial^2 F}{\partial V^2}\right|_V$ is the isothermal bulk modulus. A similar relation connects the QP correction $\Delta\alpha_V^\mathrm{QP}$ to the QP Gr\"uneisen parameter shift $\Delta\gamma$ \cite{Allen2020}
\begin{equation*}
    \Delta\alpha_V^\mathrm{QP}(T)=\dfrac{C_V}{B_TV}\Delta\gamma,
\end{equation*}
where $\Delta\gamma$ is defined by
\begin{equation*}
    \Delta\gamma=-\dfrac{V}{\omega}\left.\dfrac{\partial\Delta\omega^{(2)}}{\partial V}\right|_T.
\end{equation*}
The frequency shift $\Delta\omega^{(2)}$ arises from the lowest cubic and quartic terms in the anharmonic Halmitonian. They only entail sums of Bose-Einstein occupation functions $n$, whereas higher order terms (fifth and above) involve products \cite{Maradudin1962,Cowley1968}. At high temperatures ($k_BT\ll\hbar\omega)$, $n$ varies almost linearly with temperature, and therefore so does $\Delta\omega^{(2)}$. Consequently, the frequency shift can be identified with the first term in the temperature expansion of the QP frequency
\begin{align*}\label{QPshift}
    \Delta\omega^{(2)}=&\omega(V,T)-\omega(V)\approx \left.\dfrac{\partial\omega(V,T)}{\partial T}\right|_T T\nonumber\\
\approx& -\dfrac{\alpha_V}{2}T\omega(V,T). 
\end{align*}
Combining this equation with the definition of $\Delta\gamma$ gives
\begin{equation*}
    \Delta\gamma=-\dfrac{\alpha_V}{2} T (\gamma+\Delta\gamma), 
\end{equation*}
and
\begin{equation*}
    \alpha_V=\dfrac{\alpha_V^\mathrm{QH}}{1+\dfrac{\alpha_V}{2}T},
\end{equation*}
which has a unique positive solution given by
\begin{equation}\label{alpha_V_QP}
    \alpha_V^\mathrm{QP}=\dfrac{-1+\sqrt{1+2\alpha_V^\mathrm{QH}T}}{T}.
\end{equation}

\section{Determining $\omega_F$ and $\gamma_G$ using the YOCP model}\label{sec3}

All of the previously derived expressions for the isobaric and isochoric heat capacities, as well as the vibration pressure and bulk modulus in the framework of the liquid-phonon theory only use a restricted number of parameters, that are the Debye $\theta_D$ temperature, the $\theta_F$ one (related to the Frenkel relaxation time $\tau=G_\infty/\eta$), and the Gr\"uneisen constant $\gamma_G$. The Debye temperature was measured under normal conditions for most elements in the periodic table (see \cite{Ho1974}, page 10). Knowing the value of the Debye temperature and the Gr\"uneisen parameter at a reference density $\rho_0$ allows one to calculate the Debye temperature at a nearby density $\rho$ using the definition of the Gr\"uneisen parameter [Eq.~(\ref{gammaG_def})]. The Debye temperature can be obtained numerically for any density through successive steps thanks to
\begin{equation*}
\theta_D(\rho)=\theta_D(\rho_0)\left(\dfrac{\rho}{\rho_0}\right)^{\gamma_G(\rho_0)}.
\end{equation*}
The YOCP model can be used for the Gr\"uneisen parameter $\gamma_G$, as well as for the instantaneous shear modulus $G_\infty$ and the shear viscosity $\eta$ whose ratio yields $\theta_F=\hbar\omega_F/k_B$.

The Yukawa One Component Plasma (YOCP) (see, for instance \cite{Hubbard1971,Baus1980}) is a simple, versatile model that has been used to study a wide variety of systems, from non-interacting gases to dense liquids and crystalline solids. The YOCP is made of identical classical point particles embedded in a uniform neutralizing background. Its flexibility is due to the addition of an $e^{-r/\lambda}$ screening function to the pairwise Coulomb interaction potential in $1/r$. YOCP enables us to explore the whole range of more or less softened potentials, from the long-range Coulomb interactions of the one-component plasma (OCP) for infinite screening length $\lambda$ to the ultra-short range interactions of the hard sphere for $\lambda=0$. Although it is a simple pair interaction potential, purely repulsive and divergent at the origin, its variable smoothness is one of the reasons why YOCP is still an active field of research in statistical mechanics.

\begin{figure}[!ht]
    \centering
    \includegraphics[width=0.75\linewidth]{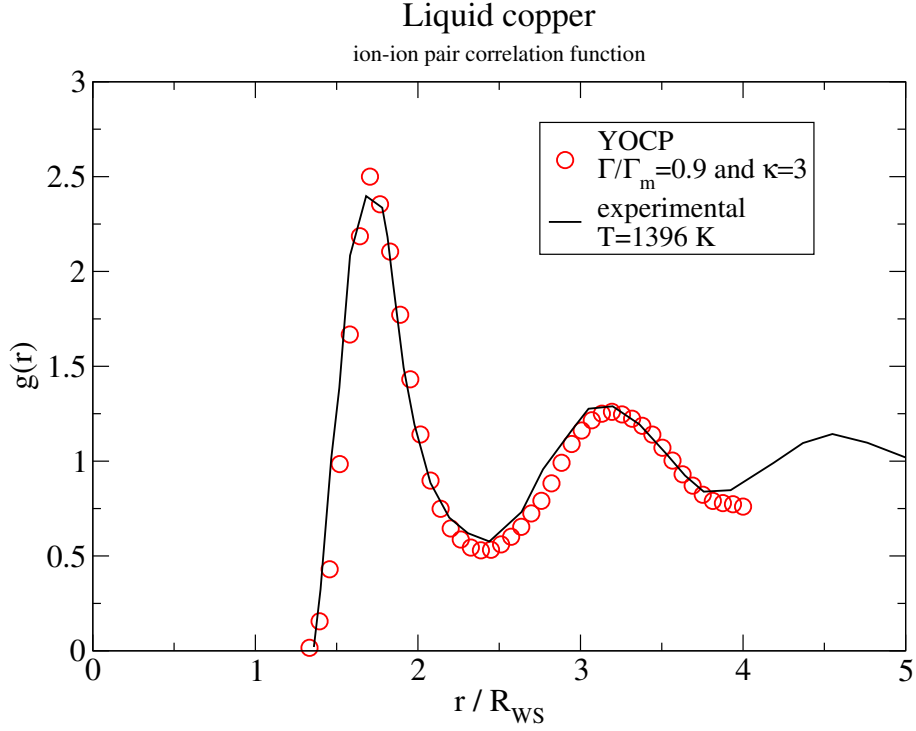}
    \caption{Liquid copper ion-ion correlation function. Black curve: experimental $g(r)$, measured at $T=1396$ K \cite{Eder1980}. red circles: YOCP $g(r)$, with parameters $\Gamma=0.9~\Gamma_m$ and $\kappa=3$, consistent with experimental density and temperature \cite{Castello2021}.}
    \label{fig:g_r}
\end{figure}

The application of the YOCP to real fluids has undergone a significant development with its identification as a Roskilde-simple (R-simple) liquid \cite{Veldhorst2015} . A R-simple liquid \cite{Ingebrigtsen2012} is a liquid model that exhibits virial and potential energy correlation of at least 90\% in its thermal equilibrium fluctuations in the NVT ensemble. R-simple liquids have isomorphic curves \cite{Gnan2009}, \emph{i.e.} curves in the phase diagram along which a large set of structural and dynamic properties remain nearly constant when expressed in properly reduced units.  These properties include the excess entropy, isochoric heat capacity, reduced-unit static and dynamic correlation functions, and reduced transport coefficients, that can be expressed as scaling laws for R-simple liquids.
An important step forward in liquid modeling is the observation that the dynamic properties are very similar for a number of systems with quite different pair potentials \cite{Bacher2014}. Systems with such non-trivial similarities are referred to as ``quasi-universal''. This follows the early finding by Rosenfeld  that the diffusion constant is an almost universal function of the excess entropy \cite{Rosenfeld1977}. It has been shown that quasi-universality applies to systems whose  pair potential can be written approximately as a sum of exponential terms with numerically large prefactors. The YOCP model belongs to this exponential class \cite{Veldhorst2015}. This supports the reliability of the YOCP model in describing properties related to ionic correlations, such as the infinite shear modulus and dynamic viscosity. Figure \ref{fig:g_r} illustrates the YOCP model’s ability to reproduce actual ion-ion correlation functions. The black line represents an experimental $g(r)$ for liquid copper \cite{Eder1980}, closely matched by the circles, which were obtained by the YOCP under similar conditions \cite{Castello2021}.

The YOCP states are specified in terms of two dimensionless variables, \emph{i.e.} the ion-ion coupling parameter $\Gamma$ and the screening parameter $\kappa$, reading respectively
\begin{equation}\label{Gamma}
    \Gamma=\dfrac{e^2}{4\pi\epsilon_0}\dfrac{{Z^*}^2}{k_BT\,R_\mathrm{ws}},
\end{equation}
and
\begin{equation*}
    \kappa=\dfrac{R_\mathrm{ws}}{\lambda},
\end{equation*}
where $R_\mathrm{WS}$ denotes the Wigner-Seitz radius, given by
\begin{equation*}
    R_\mathrm{WS} = \left(\dfrac{3n_i}{4\pi}\right)^{1/3}, 
\end{equation*}
with $n_i$ denoting the ion number density. 

 Khrapak \emph{et al.} developed a set of simple, compact parameterizations \cite{Khrapak2015,Khrapak2017,Khrapak2019,Khrapak2020,Khrapak2024} based on universal scaling laws for the quantities required by the formalism presented in the preceding section of this paper. These parameterizations agree well with values derived from molecular dynamics, Monte Carlo, and integral equation methods \cite{Gilles2007,Salin2003,Hamaguchi1997,Castello2021_int_eq} across much of the ($\Gamma$,$\kappa$) phase diagram.

The next two sections respectively specify the expressions for the screening length $\lambda$ and for the ionization $Z^*$, necessary to obtain the YOCP variables $\Gamma$ and $\kappa$. The subsequent sections present the recent parameterizations by Khrapak \emph{et al.}  useful for the calculation of $\omega_F$ and $\gamma_G$. 

\subsection{Thomas-Fermi ionization}\label{subsec31}

We propose to use the fit of the Thomas-Fermi ionization published by More \cite{More1985}
\begin{equation}\label{Z*_More}
    Z^*=Z \dfrac{x}{1+x+\sqrt{1+2x}},
\end{equation}
where $Z$ is the atomic number, and $x$ is given by
\begin{equation*}
    x=14.3139~(R^C+Q_1^C)^{0.6624/C}.
\end{equation*}
In this expression, $R$ is defined as
\begin{equation*}
    R=\dfrac{\rho}{ZA},
\end{equation*}
$A$ denoting the molar mass, in grams, and the density $\rho$ is in g.cm$^{-3}$. $Q_1$ is the quantity
\begin{equation*}
    Q_1=(0.3323\times 10^{-2}~T_0^{0.9718}+0.926148\times 10^{-4}~T_0^{3.10165})\nonumber\\
               \times R^{-\exp(-1.763+1.43175~T_\mathrm{F}+0.315463~T_\mathrm{F}^7)}, 
\end{equation*}
where
\begin{equation*}
    T_0=\dfrac{T_\mathrm{eV}}{Z^{4/3}},
\end{equation*}
$T_\mathrm{eV}$ being the temperature in eV, and $T_F$ the ratio
\begin{equation*}
    T_F=\dfrac{T_0}{1+T_0}.
\end{equation*}
Finally, the quantity $C$ is given by
\begin{equation*}
    C=-0.366667\,T_F +0.983333. 
\end{equation*}

\subsection{Thomas-Fermi screening parameter $\kappa$}\label{subsec32}

We will use the finite-temperature Thomas-Fermi screening length for the calculation of the screening parameter $\kappa$, which reads \cite{Gilles2007}
\begin{equation}\label{kappa}
    \kappa=\dfrac{R_\mathrm{ws}}{\lambda_\mathrm{TF}}=\dfrac{(2 r_s Z^*)^{3/4}}{\Gamma^{1/4}}\dfrac{\left[I_{-1/2}\left(\dfrac{\mu}{k_BT}\right) \right]^{1/2}}{\sqrt\pi},
\end{equation}
where $\mu$ denotes the chemical potential and $r_s$ is the electron correlation parameter
\begin{equation*}
    r_s=\dfrac{1}{a_B}\left(\dfrac{3}{4\pi n_e} \right)^{1/3},
\end{equation*}
$a_B$ being the Bohr radius and $n_e=Z^* n_i$ the electron number density. Alternatively, $r_s$ reads $r_s=\dfrac{1}{a_B}\dfrac{R_\mathrm{ws}}{{Z^*}^{1/3}}$. $I_{-1/2}$ is the Fermi integral of order $-1/2$
\begin{equation*}
    I_{-1/2}(x)=\int_0^\infty\,\mathrm{d}t \dfrac{t^{-1/2}}{e^{(t-x)}+1},
\end{equation*}
and $x=\dfrac{\mu}{k_BT}$ is the solution of
\begin{equation}\label{inversion}
    I_{1/2}(x)=\dfrac{\pi^2}{\sqrt{2}}\left(\dfrac{\hbar^2}{m_e}\right)^{3/2} (k_BT)^{-3/2}n_e,
\end{equation}
where $I_{1/2}$ is the Fermi integral of order $1/2$
\begin{equation*}
    I_{1/2}(x)=\int_0^\infty\,\mathrm{d}t \dfrac{t^{1/2}}{e^{(t-x)}+1}.
\end{equation*}
The solution of Eq.~(\ref{inversion}) has been parameterized by Karasiev, Chakraborty and Trickey \cite{Karasiev2015}.

Within our study of the liquid state where $\dfrac{k_BT}{\epsilon_F}\ll 1$, one has
\begin{equation*}
    \dfrac{\mu}{k_BT}\approx \dfrac{\epsilon_F}{k_BT} \left[1-\dfrac{\pi^2}{12}\left(\dfrac{k_BT}{\epsilon_F} \right) ^2 \right].
\end{equation*}
Using the asymptotic expression of the Fermi integral $I_{-1/2}$
\begin{equation*}
    I_{-1/2}(x)\approx 2 x^{1/2}-\dfrac{\pi^2}{12}x^{-3/2}
\end{equation*}
yields the following approximation for the screening parameter $\kappa$
\begin{align*}
    \kappa\approx \dfrac{2}{\sqrt{\pi}}\left(\dfrac{9\pi}{4}Z^*\right)^{1/6} \left(\dfrac{R_\mathrm{WS}}{a_B}\right)^{1/2}+O\left(\dfrac{k_BT}{\epsilon_F}\right)^2. 
\end{align*}

\subsection{The frequency $\omega_F$ using the YOCP model}\label{subsec33}

Khrapak studied the dependence of both $G_\infty$ and $\eta$ across extended ranges of thermodynamic conditions to the fluid model (YOCP, Lennard-Jones [LJ], soft-spheres [SS] and hard-spheres [HS]). In Ref.~\cite{Khrapak2024quasiuniversal}, the author underlines the quasi universal behavior of the phonon relaxation time when it is expressed in properly reduced units, although both normalized quantities $\overline{G}_\infty$ and $\overline{\eta}$ are sensitive to the fluid model. The reduced relaxation time $\overline{\tau}=1/\overline{\omega}_F=\overline{\eta}/\overline{G}_\infty$ obtained with the four models are found to exhibit very close behaviors when the fluid's density evolves from the gaz-like limit to the dense liquid near melting. In particular, the values of $\overline{\tau}$ observed at its minimum and at melting are found to be very close. We therefore expect the parameterizations of $\overline{G}_\infty$ and $\overline{\eta}$ from the YOCP (as well as those from the LJ, SS and HS systems) to give relevant values for $\overline{\tau}$.

\subsubsection{The YOCP instantaneous shear modulus}

The YOCP excess instantaneous shear modulus is related to the radial distribution function $g(r)$ by \cite{Khrapak2020}
\begin{equation*}
    \Delta G_\infty=\dfrac{m_i n_i \omega_p^2 R_\mathrm{ws}^2}{30} \int_0^\infty\,x\,g(x)\,e^{-\kappa x}(\kappa^2 x^2-2\kappa x-2)\,\mathrm{d}x, 
\end{equation*}
where $x=r/R_\mathrm{ws}$. $m_i$ is the ion mass, and the square of the plasma frequency is given by
\begin{equation*}\label{omega_p}
    \omega_p^2=\dfrac{e^2}{\epsilon_0}\dfrac{{Z^*}^2 n_i}{m_i}.
\end{equation*}
Khrapak and Klumov calculated $\Delta G_\infty$ using molecular dynamics radial distribution functions, for screening parameters $\kappa=1,2,3$ and 4, and parameterized with a good precision their results as  \cite{Khrapak2020}
\begin{equation*}
    \dfrac{\Delta G_\infty}{m_i n_i\omega_p^2 R_\mathrm{ws}^2}=\dfrac{\kappa^4 [(\kappa^2+3)\sinh(\kappa)-3\kappa\cosh(\kappa)]}{45[\kappa\cosh(\kappa)-\sinh(\kappa)]^3}. 
\end{equation*}
Using the equality $m_i n_i \omega_p^2 R_\mathrm{ws}^2=3\Gamma\, n_i k_BT$, the total YOCP instantaneous shear modulus $G_\infty$ is given by
\begin{equation}\label{Khrapak_Ginf}
    G_\infty=\overline{G}_\infty\,n_i k_BT,
\end{equation}
with
\begin{equation*}
    \overline{G}_\infty=1+\dfrac{3\Gamma}{45}f(\kappa), 
\end{equation*}
where we have introduced the function
\begin{equation*}
    f(\kappa)=\dfrac{\kappa^4 [(\kappa^2+3)\sinh(\kappa)-3\kappa\cosh(\kappa)]}{[\kappa\cosh(\kappa)-\sinh(\kappa)]^3}. 
\end{equation*}

\subsubsection{The YOCP shear viscosity}\label{subsec44}

Khrapak \cite{Khrapak2024} parameterized the YOCP (with $\kappa=1,2$ and 3, and $2 \leq\Gamma\leq 1000$) molecular dynamics shear viscosities of Donko and Hartmann \cite{Donko2008}, and the OCP ones ($2\leq\Gamma\leq 402$) of Daligault, Rasmussen and Balrud \cite{Daligault2014}, following Rosenfeld's scaling relation between transport coefficients and excess entropy \cite{Rosenfeld1977}, together with a modified Rosenfeld-Tarazona scaling \cite{Rosenfeld1998} of the excess entropy.

Khrapak's parameterization reads \cite{Khrapak2024}
\begin{equation}\label{Khrapak_eta}
    \eta=\overline{\eta}\ m_i v_T n_i^{2/3},
\end{equation}
where $v_T=\sqrt{k_BT/m_i}$ is the thermal velocity.
The reduced viscosity $\overline{\eta}$ is given by
\begin{equation*}
    \overline{\eta} = 0.13\,e^{-0.9s_\mathrm{ex}}, 
\end{equation*}
where $s_\mathrm{ex}$ denotes the reduced excess entropy, for which the author proposes a modified Rosenfeld-Tarazona scaling with the reduced coupling strength $\Gamma/\Gamma_m$, $\Gamma_m$ being the ion-ion coupling parameter at melting.  The $2/5$ exponent of $\Gamma/\Gamma_m$ is replaced by $1/2$ yielding
\begin{equation*}
    s_\mathrm{ex}=s_m\left(\dfrac{\Gamma}{\Gamma_m} \right)^{1/2}, 
\end{equation*}
where $s_m$ is the reduced excess entropy at melting, given by
\begin{equation*}
    s_m=-4.109+0.096~\kappa.
\end{equation*}
This modification of the original Rosenfeld-Tarazona scaling relation was found by Khrapak to improve significantly the agreement with molecular dynamics simulations.  
The YOCP ion-ion coupling parameter $\Gamma_m$ at melting is expressed as \cite{Khrapak2015}
\begin{equation*}
    \Gamma_m=172\,\dfrac{e^{\alpha\kappa}}{1+\alpha\kappa+(\alpha\kappa)^2/2},
\end{equation*}
where $\alpha=\left(4\pi/3 \right)^{1/3}$.

\subsection{YOCP Gr\"uneisen parameter and thermal expansion parameter}\label{subsec34}

Due to the strong virial-potential (W-U) correlations that characterize R-simple liquids, temperature is a linear function of density on a log-log scale along isomorphs for these systems. This defines the density exponent:
\begin{equation*}
\gamma\equiv \left.\dfrac{\partial\ln T}{\partial\ln\rho}\right|_{s_\mathrm{ex}},   
\end{equation*}
which is related to virial and potential fluctuations $\Delta W$ and $\Delta U$ by:
\begin{equation*}
\gamma=\dfrac{\langle \Delta W \Delta U \rangle}{\langle(\Delta U)^2\rangle}\approx \dfrac{\left.\partial W/\partial T \right|_V}{\left.\partial U/\partial T \right|_V}.    
\end{equation*}
The second equality is a consequence of the high W-U correlation characteristic of R-simple liquids, that allows to identify $\gamma$ with the linear regression slope through the $\Delta W-\Delta U$ thermal fluctuations data. The density exponent $\gamma$ given by this slope is related to the R-simple liquid's Gr\"uneisen parameter $\gamma_G$ by \cite{Schroder2009}:
\begin{equation*}\label{Schroder_gruneisen}
\gamma_G=\dfrac{\gamma~\overline{C}_V^\mathrm{ex}+1}{\overline{C}_V},
\end{equation*}
where $\overline{C}_V=C_V/Nk_B$ is the reduced heat capacity, and $\overline{C}_V^\mathrm{ex}=C_V^\mathrm{ex}/Nk_B$ the reduced excess heat capacity. Applying the Rosenfeld-Tarazona scaling, the excess contribution due to ion-ion coupling reads:
\begin{equation*}\label{Khrapak_CV}
    \overline{C}_V \approx \dfrac{3}{2}+\dfrac{3\delta}{5} \left(\dfrac{\Gamma}{\Gamma_m} \right)^{2/5},
\end{equation*}
where Khrapak recommends to take $\delta=3.1$.

Khrapak developed the following expression for the density exponent $\gamma$ of the Yukawa fluid, by deriving expressions of $W$ and $U$:
\begin{equation*}
    \gamma(x)=\dfrac{1}{3}\dfrac{(2+2x+x^2+x^3)}{(2+2x+x^2)}.
\end{equation*}
The potential energy $U$ and the virial $W$ are expressed as sums of static Ewald and thermal contributions. Dynamic effects due to the persistence of transient shear phonon modes, are not accounted for. Due to significant discrepancies between the calculated and measured parameters in the immediate vicinity of the melting point, we opt to use the approximate form proposed by Khrapak: 
\begin{equation}\label{Khrapak_gruneisen}
    \gamma_G\approx\dfrac{2}{3} + \dfrac{2}{15}\delta \left[3\gamma(\alpha\kappa)-2 \right] \left(\dfrac{\Gamma}{\Gamma_m} \right)^{2/5}. 
\end{equation}
This expression serves as a reliable approximation for warm and moderately coupled liquids ($\Gamma/\Gamma_m<0.5$), for which the Khrapak's expression for the density exponent parameter $\gamma$ becomes relevant. Although its extension to the vicinity of the melting is questionable, we found that this choice appreciably reduces discrepancies between the YOCP $\gamma_G$ and values deduced from shock/release experiments at melting (in the absence of theoretical or experimental values for the density exponent parameter $\gamma$). 

The product $\gamma_G \overline{C}_V$ yields the numerator of the YOCP thermal expansion parameter
\begin{equation*}\label{alpha_V_YOCP}
    \alpha_V=\dfrac{\gamma_G C_V}{B_T V}=\dfrac{\gamma_G\overline{C_V}}{\mu T}.
\end{equation*}
$\mu$ denotes the reduced isothermal bulk modulus, defined by the ratio of the physical isothermal bulk modulus $B_T$ and $Nk_BT/V$, and parameterized as follows \cite{Khrapak2015}:
\begin{equation*}
    \mu(\kappa,\Gamma)=\dfrac{B_T}{Nk_BT/V}=\left(1+\dfrac{\epsilon}{3}\right)+\dfrac{\Gamma \kappa^6 \sinh\kappa}{9\left[ \kappa\cosh\kappa-\sinh\kappa\right]^3}\\
                       +\dfrac{\delta}{45}\left(\dfrac{\Gamma}{\Gamma_m} \right)^{2/5} f_\mu(\alpha\kappa). 
\end{equation*}
Khrapak recommends to take $\epsilon=-0.1$ and $\delta=3.2$. We recall that $\alpha=\left(4\pi/3\right)^{1/3}$. The function $f_\mu(x)$ is defined by
\begin{equation*}    f_\mu(x)=\dfrac{2x^6+14x^5+35x^4+76x^3+136x^2+136x+68}{(x^2+2x+2)^2}.
\end{equation*}

We express the same reservations regarding the expression for the reduced bulk modulus $\mu$ as we do for the parameterization of the Gr\"uneisen coefficient, concerning its validity in the vicinity of the melting curve, where viscosity effects may be significant. The calculation of $\mu$ in this region may be very approximate, but becomes reasonably realistic as the liquid becomes more fluid. In the next section, we evaluate this formalism by applying it to the calculation of the specific heat at constant pressure for two simple metals in the liquid state (copper and aluminum).

\section{Interpretation of isobaric heat capacity $C_P(T)$ measurements for aluminum and copper}\label{sec4}

\subsection{Available experimental data}\label{subsec41}

In this section we examine the isobaric heat capacities derived from Gathers's experiments on aluminum and copper performed at the isobaric expansion apparatus IEX in Livermore at a pressure of 0.3 GPa \cite{Gathers1983}. The author measured the specific enthalpies with 4\% precision and the specific volumes with 2\% and 6\% accuracy for copper and aluminum, respectively. The temperatures were inferred from the enthalpies using the 1973 Hultgren tables with a precision estimated at 4\%. 

The frequency $\omega_F$ plays a crucial role in liquid-phonon theory. It represents the increase in fluidity (the inverse of the viscosity $\eta$) that occurs at the expense of shear resistance as the liquid's temperature rises. Experimental viscosities are available for liquid copper and aluminum, but only within restricted ranges of temperature above melting. In liquids, the static (\emph{i.e.} the zero-frequency) shear modulus vanishes. Only resistance to high-frequency shear stress load persists. The infinite-frequency shear modulus $G_\infty$ is an idealized quantity, \emph{a priori} inaccessible by experimentation due to limitations in the ability to produce the extremely high frequencies required. However, Puosi and Leporini demonstrated through computer simulations that the relevant high-frequency shear modulus controlling relaxation corresponds to shear modulus values measured over timescales that, while much shorter than any relaxation time, are still significantly longer than typical vibration periods \cite{Puosi2012}. The relaxation time of viscous liquids exhibits a non-Arrhenius temperature dependence, which, according to the so-called shoving model \cite{Dyre1996,Dyre1998}, is influenced by the idealized infinite-frequency shear modulus. Dyre and Wang compared data for a large selection of metallic glasses to three different versions of the shoving model and established the correspondence between $G_\infty$ and the experimentally accessible high-frequency plateau modulus \cite{Dyre2012}. 

We could not find any experimental high-frequency values for the pure copper and aluminum liquids. However, the fluid's response to sudden (high-frequency) perturbations is not significantly different from that of a solid body \cite{Zwanzig1965}. Furthermore, dynamic and static moduli are comparable for solids \cite{Ledbetter1993}. Indeed, dynamic measurements are adiabatic, while static ones are isothermal. The isothermal and adiabatic shear moduli are identical at all temperatures since deformations at constant volume are adiabatic if isothermal \cite{Ledbetter1993}.
Static shear modulus measurements for metallic solids up to their melting temperature, when combined with viscosity measurements, can be used to evaluate the $\omega_F$ frequencies in the liquid near the melting point. The shear modulus is an average of two shear moduli, which are related to the elastic constants of the crystal. For cubic crystals, these are $C=C_{44}$ and $C^\prime=(C_{11}-C_{12})/2$ known as rhombohedral and tetragonal shear moduli respectively. According to the Born criterion, a crystal is unstable if at least one of $C$ and $C^\prime$ vanishes. Using QMD simulations, Wang \emph{et al.} examined how the two cubic shear moduli vary with temperature up to the melting point, and concluded that it is the tetragonal shear modulus $C^\prime$ that disappears at melting rather than the rhombohedral shear modulus $C$ \cite{Wang1997}. Rather than extrapolating the experimental values of $G(T)$ to the melting point, it is preferable to extrapolate from values of $C=C_{44}$ when available. This is possible for copper and aluminum \cite{Chang1966,Gerlich1969}.

Both Gr\"uneisen parameter $\gamma_G$ and thermal expansion parameter $\alpha_V$ play an important role in the $C_P(T)$ curves. $\gamma_G$ is not directly measurable. It can be inferred from neutron or X-ray inelastic scattering experiments, or Raman spectroscopy under pressure, this latter method investigating Gr\"uneisen constants for optical phonon modes. A current method consists in fitting models for $\gamma_G$ on shock/release experiments via an equation of state. Values obtained in this way are available for the Gr\"uneisen parameters of liquid copper \cite{Gilev2018} and liquid aluminum \cite{Gilev2020} at their melting temperatures. The thermal expansion parameter is also not directly measurable. Gathers provides the values of the specific volumes of liquid copper and liquid aluminum over the temperature range studied \cite{Gathers1983}. We compare them to the $V(T)$ curves obtained by integrating the YOCP $\alpha_V(T)$.

The analysis undertaken in the following section aims to address the following points:
\begin{itemize}
    \item The relevance of the liquid-phonon approach: is the finite lifetime of transverse phonon modes clearly reflected in the thermodynamic properties? Does taking this into account result in a marked change in the behavior of the isobaric specific heat $C_p(T)$?
    \item We suspect that the parametrization of the YOCP Gr\"uneisen parameter $\gamma_G$ and of the YOCP bulk modulus $\mu$ are suboptimal for liquids close to the melting curve. How do these uncertainties affect our theoretical $C_P(T)$ values? 
    \item We have introduced anharmonicity at the lowest order in the formalism. To what extent is this approximation still acceptable?
\end{itemize}

\subsection{Liquid copper}\label{subsec42}

Gathers fitted a quadratic polynomial to his experimental enthalpies $H$ in the range 2000 K $\leq T\leq$ 4500 K \emph{excluding the melting point}, and obtained
\begin{equation*}
    H[\mathrm{MJ.kg}^{-1}]=-0.22367+6.8142\times 10^{-4}~T\nonumber\\
    -3.1631\times 10^{-8}~T^2.
\end{equation*}
The evolution of the specific isobaric heat capacity $C_P$ with temperature (in the range 2000 K $\leq T\leq$ 4500 K) follows straightforwardly; one has
\begin{equation}\label{Cp_Gathers}
    C_P[\mathrm{J.g^{-1}.T^{-1}}]=0.68142-6.3262\times 10^{-5}~T.
\end{equation}
We note that the $T^2$ term in the expression that fits the experimental enthalpies of liquid copper contributes significantly, reaching up to 16\% (\emph{i.e.} four times the uncertainty on the experimental enthalpies) of the value of the linear term, reinforcing our confidence in the temperature dependence of $C_P(T)$.
The persistence of a partial local order in copper in the liquid state, which gradually disappears with increasing temperature, is obvious from the negative slope. According to the conventional model of the liquid as a purely viscous medium, $C_V$ should be constant (equal to $2Nk_B$), and $C_P$ should increase with temperature $T$. The negative slope observed in the experimental $C_P$ curve reflects the compensation of the thermal expansion, which is responsible for the usual temperature dependence, by the progressive decrease of $C_V$, from $3Nk_B$ to $2Nk_B$, which is induced by the progressive loss of order caused by the decrease of the viscosity. The question remains as to whether the YOCP model can best reproduce these experiments. To this end, we examine the main stages of the calculations and compare the YOCP quantities with the available experimental data.

In Table~\ref{tab:copper1}, we present the ion charges $Z^*$, the screening parameter $\kappa$, and the ion coupling constant $\Gamma$ calculated for the temperatures and densities given in the two first columns. They are respectively given by Eqs.~(\ref{Z*_More}), (\ref{kappa}) and (\ref{Gamma}).
\begin{table}[!ht]
    \centering
    \begin{tabular}{c c c c c c}
    \hline\hline
     $T$ (K) & $\rho$ (g.cm$^{-3}$) & $Z^*$ & $\kappa$ & $\Gamma$ \\
     \hline
    1350 & 7.832 & 4.112 & \hspace{0.15cm}3.306\hspace{0.15cm} & 1411 \\
    2000 & 7.516 & 4.032 & 3.317 & 907 \\
    2250 & 7.319 & 3.981 & 3.325 & 779 \\
    2500 & 7.086 & 3.918 & 3.334 & 672 \\
    2750 & 6.921 & 3.874 & 3.341 & 592 \\
    3000 & 6.764 & 3.831 & 3.347 & 527 \\
    3250 & 6.565 & 3.774 & 3.356 & 468 \\
    3500 & 6.423 & 3.735 & 3.362 & 422 \\
    \hline\hline
    \end{tabular}
    \caption{Liquid copper: the two first columns report the temperatures and densities of copper in the experiments of Gathers \cite{Gathers1983}. The third column gives the corresponding ion charges $Z^*$ (Eq.~(\ref{Z*_More})), and the fourth and fifth, respectively the corresponding YOCP parameters $\kappa$ and $\Gamma$ [Eqs.~(\ref{kappa}) and (\ref{Gamma})].
    }
    \label{tab:copper1}
\end{table}
These values are used to calculate the Debye temperature $\theta_D$, the temperature $\theta_F$ [using Eqs.~(\ref{Khrapak_Ginf}) and (\ref{Khrapak_eta})], the Gr\"uneisen constant $\gamma_G$ [Eq.~(\ref{Khrapak_gruneisen})], and the expansion coefficient $\alpha_V$ [Eq.~(\ref{alpha_V_QP})] which are needed to obtain the isobaric heat capacities $C_P(T)$. They are given in Table~\ref{tab:copper2}. 

\begin{table}[!ht]
    \centering
    \begin{tabular}{c c c c c}
    \hline\hline
     $T$ (K) & $\theta_D$ (K) & $\theta_F$ (K) & $\gamma_G$ & $\alpha_V$ (K$^{-1}$) \\
     \hline
    1350 & 252 & 82 &\hspace{0.15cm} 1.70 \hspace{0.15cm} & 1.02$\times 10^{-4}$ \\
    2000 & 236 & 124 & 1.53 & 8.41$\times 10^{-5}$ \\
    2250 & 227 & 138 & 1.48 & 7.97$\times 10^{-5}$ \\
    2500 & 217 & 149 & 1.43 & 7.64$\times 10^{-5}$ \\
    2750 & 210 & 159 & 1.39 & 7.31$\times 10^{-5}$ \\
    3000 & 203 & 167 & 1.36 & 7.03$\times 10^{-5}$ \\
    3250 & 195 & 174 & 1.33 & 6.81$\times 10^{-5}$ \\
    3500 & 190 & 180 & 1.30 & 6.59$\times 10^{-5}$ \\
    \hline\hline
    \end{tabular}
    \caption{Liquid copper: the table gives the values of the Debye temperature $\theta_D$, the temperature $\theta_F=\hbar\omega_F/k_B$, the YOCP Gr\"uneisen constant $\gamma_G$ and the YOCP thermal expansion parameter $\alpha_V$ (including QP corrections), corresponding to the screening and ion coupling parameters of Table \ref{tab:copper1}.}
    \label{tab:copper2}
\end{table}

\subsubsection{Assessment of the accuracy of $\omega_F$, $\gamma_G$ and $\alpha_V(T)$ obtained using the YOCP model.}

\begin{figure}[!ht]
    \centering
    \includegraphics[width=0.75\linewidth]{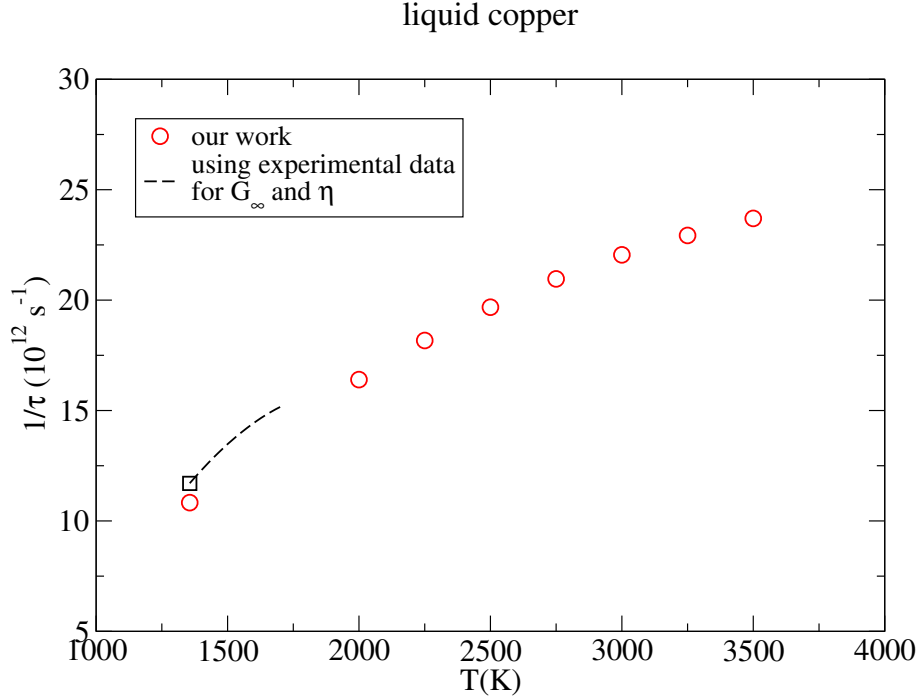}
    \caption{The figure shows the YOCP ratio $G_\infty/\eta$ (red circles) and compares the values at temperatures close to melting to estimations (black square and dashed line) deduced from the available experimental data (Chang and Himmel for $G_\infty$ \cite{Chang1966} and Assael \emph{et al.}'s fit of experimental shear viscosity between 1356 K and 1970 K \cite{Assael2010}). 
    }
    \label{fig:copper5}
\end{figure}
Figure~\ref{fig:copper5} shows the evolution of the YOCP $1/\tau=G_\infty/\eta$ ratio (red circles). To assess the relevance of these calculations, we present an estimate of this ratio near the melting point derived from experimental data. 
Assael \emph{et al.} compiled available experimental data for the density and viscosity of liquid copper, in order to establish standards for these quantities \cite{Assael2010}. Collected viscosity data for temperatures 1356 K $\leq T\leq$ 1970 K, meeting eight quality criteria, are fitted by the expression
\begin{equation*}
    \log_{10}\left(\dfrac{\eta}{\eta_0}\right)=-a_1+\dfrac{a_2}{T},
\end{equation*}
with $\eta_0=10^{-3}$ Pa.s, $a_1=0.4220$ and $a_2=1393.4$ K. The standard deviation is 6.3\%, with a confidence level of 95\%. We estimate the value of $G_\infty$ in the liquid state near the melting point using $C_{44}$ measurements. Chang and Himmel's values \cite{Chang1966} can be fitted by the following linear relation between 300 K and 800 K:
\begin{equation*}
C_{44}=83.965-2.7109\times 10^{-2}\,T.
\end{equation*}
By extrapolating this to the melting temperature $T_m=1356$ K, we obtain $G_\infty\equiv C_{44}\approx 47.2$ GPa. The black square shows the ratio of this value to the viscosity, which is $\eta=4$ mPa.s, as given by the formula of Assael \emph{et al.} at the melting temperature. The YOCP value at the melting point is fairly close to the experimental value $\omega_F=1/\tau$ represented by the square. We extrapolated the experimental curve $C_{44}(T)$ a little further to propose a possible temperature dependence of $1/\tau$ based on experiments. This is represented by the black dashed line. There is rather good agreement with the tendency followed by our calculations (red circles) based on the YOCP. We therefore conclude that the YOCP theory provides an accurate estimate of the value of $\omega_F$, at least up to 600 K above the melting temperature. Conversely, our confidence in the high-temperature YOCP estimates is based on the fact that $\omega_F/\omega_D$ gradually approaches 1 without exceeding this upper limit (see Table \ref{tab:copper2}).
 
\begin{figure}[!ht]
    \centering
    \includegraphics[width=0.75\linewidth]{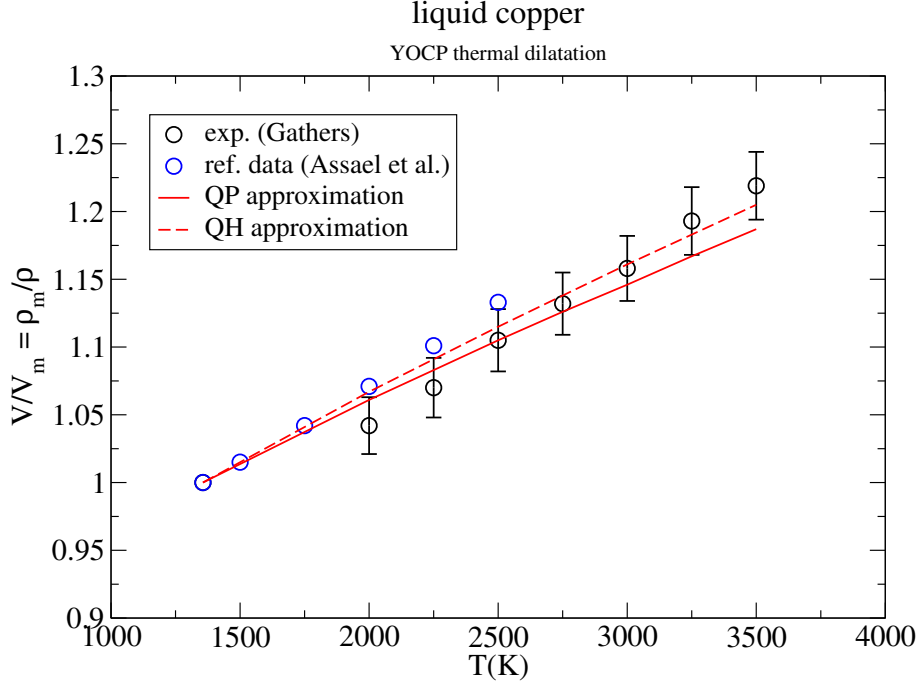}
    \caption{Liquid copper volume expansion at constant pressure. Black circles: experimental values of Gathers \cite{Gathers1983}. Error bars: 2\%. Blue circles: reference data of Assael \emph{et al.} \cite{Assael2010}. Red full curve: our work, where we applied quasi-particle (QP) corrections to the YOCP thermal dilatation coefficient $\alpha_V$. Red dashed curve: quasi-harmonic (QH), i.e., the same work, without QP corrections.}
    \label{fig:copper4}
\end{figure}

An estimate of the value of liquid copper's Gr\"uneisen parameter at melting temperature can be inferred from a semi-empirical expression for $\gamma_G$ given by Gilev in Ref.\cite{Gilev2018}: 
\begin{equation}\label{Gruneisen_Gilev}
\gamma_G^\mathrm{Gilev} (V,T)=\gamma_\infty+\left(\dfrac{V}{V_0}\right)^\delta \dfrac{\gamma_0-\gamma_\infty}{1+\beta T_0} (1+\beta T),
\end{equation}
where $\gamma_\infty=2/3$, $T_0$ and $V_0$ denote the reference temperature and specific volume, and $\gamma_0\equiv \gamma(V_0,T_0)$. The remaining two parameters $\delta$ and $\beta$ are fitted on shock/release curves. For liquid copper, Gilev obtains $\delta=-2$ and $\beta=1.9\times 10^{-6}$ K$^{-1}$. The estimate for liquid copper with $\rho=7.832$ g$\cdot$cm$^{-3}$ and $T=$ 1350 K is $\gamma_G^\mathrm{Gilev}=1.66$, which is close to the value $\gamma_G=1.70$ we calculate using Eq.~(\ref{Khrapak_gruneisen}).

In the previous section, we questioned the relevance of resorting to normalized YOCP heat capacity $\overline{C}_V$ and normalized bulk modulus $\mu$ to calculate the thermal expansion parameter $\alpha_V$. Figure~\ref{fig:copper4} compares the volume expansion $V(T)$ derived from the YOCP $\alpha_V$ with the experimental values of Gathers that were used as inputs in our calculations. The volumes $V(T)$ are obtained from 
\begin{equation*}
    \ln(V/V_m)=\int_{T_m}^T \alpha_V(T)\,\mathrm{d}T,
\end{equation*}
where $V_m$ denotes the volume at melting. Gathers's experimental values are represented by black empty circles with their error bars. The figure also shows Assael \emph{et al.}'s recommended values \cite{Assael2010} as blue empty circles. The two curves correspond to our evaluations derived from the YOCP thermal expansion parameters, using the QH approximation (red dashed line) or accounting for QP corrections (red full line). The YOCP expansion parameter provides rather good agreement with the experiments as well as with the QH approximation or when the QP corrections are considered. However, the latter seem to underestimate the thermal expansion at the highest temperatures, where they fall outside the $2\%$ error bar estimated by Gathers.

\subsubsection{Application to $C_P(T)$}

\begin{figure}[!ht]
    \centering
    \includegraphics[width=0.75\linewidth]{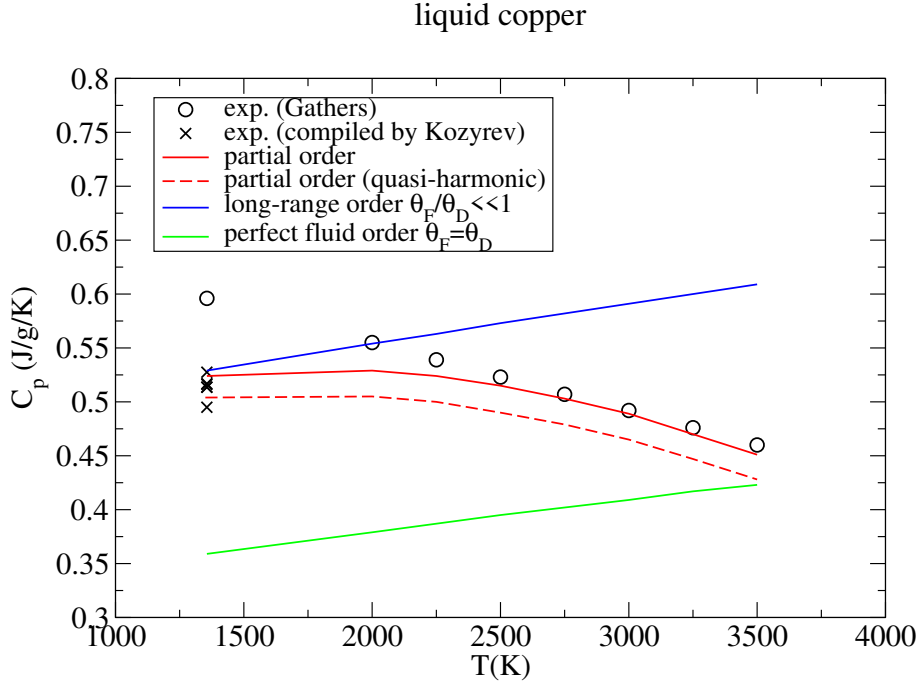}
    \caption{Liquid copper heat capacity $C_P$ at constant pressure. Black circles: experimental values of Gathers \cite{Gathers1983}. Black ``+ signs'': experimental values at melting temperature, reported in \cite{Kozyrev2023}. Red curve: our work, assuming persistence of partial-local crystalline order. Blue curve: the same work, considering persistence of long-range crystalline order at any temperature. Green curve: no crystalline order persists in the liquid. All of the solid lines were obtained using the quasi-particle approximation. The red dashed curve is obtained with the quasi-harmonic approximation.}
    \label{fig:copper4b}
\end{figure}

Figure~\ref{fig:copper4b} compares our calculations of the specific heat capacities at constant pressure $C_P(T)$ of liquid copper assuming three different structural orders with experimental data. Gathers's values are shown by circles. The first point at $T=1350$ K is outside the range of validity ($T\geq$ 2000 K) of the fit given by Eq.~(\ref{Cp_Gathers}). Kozirev collected other measurements performed up to 1800 K or 2900 K, depending on the source \cite{Kozyrev2023}. These measurements are represented by ``+'' signs at melting temperature. The author recommends the value $C_P = 0.527$ J.g$^{-1}$.K$^{-1}$ for $T_m\leq T\leq$ 2900 K from Ablaster's recent study \cite{Arblaster2015}.

Our calculations correspond to the three solid lines. They are also reported in Table \ref{tab:copper3}. The blue line assumes the persistence of long range crystalline order in the liquid at all temperatures. It is obtained by setting $\theta_F=0$ in the formalism. The green curve, obtained with $\theta_F=\theta_D$, describes the case of perfectly fluid liquid, whith total absence of crystalline order. The red line shows the case of partial-local-order persistence, where $\theta_F$ varies as the ratio $G_\infty/\eta$. Clearly the latter approximation best describes the temperature dependence of the constant-pressure heat capacity of liquid copper. The results represented by the red line agree with Gathers's points at $T\gtrsim$ 2500 K and with the value $C_P = 0.527$ J.g$^{-1}$.K$^{-1}$ recommended by Kozyrev below. Finally, the red dashed curve shows our calculations assuming partial order when QP corrections are neglected. Although Figure~\ref{fig:copper4} suggests that the QH approximation is preferable to the QP one, the latter significantly improves agreement with the experimental $C_P$ at high temperature.

\begin{table}[!ht]
    \centering
    \begin{tabular}{c c c c c}\hline\hline
     $T$(K) & $C_P^\mathrm{exp}$ & $C_P^\mathrm{po}$ & $C_P^\mathrm{lro}$ & $C_P^\mathrm{pfo}$ \\
    & \multicolumn{4}{c}{(J/g/K)}\\
     \hline
    1350 & \hspace{0.15cm} 0.596 \hspace{0.15cm} & \hspace{0.15cm} 0.524 \hspace{0.15cm} & \hspace{0.15cm} 0.529 \hspace{0.15cm} & \hspace{0.15cm} 0.359 \hspace{0.15cm} \\
    2000 & 0.555 & 0.528 & 0.554 & 0.379\\
    2250 & 0.539 & 0.524 & 0.563 & 0.386\\
    2500 & 0.523 & 0.515 & 0.573 & 0.394\\
    2750 & 0.507 & 0.503 & 0.582 & 0.402\\
    3000 & 0.492 & 0.489 & 0.591 & 0.409\\
    3250 & 0.476 & 0.470 & 0.600 & 0.417\\
    3500 & 0.460 & 0.451 & 0.609 & 0.424\\
    \hline\hline
    \end{tabular}
    \caption{Liquid copper: the table compares our $C_P$ calculations (columns 3,4 and 5) to the values from Gathers's experimental enthalpies (column 2). The superscripts "po", "lro", and "pfo" denote the following cases, respectively: persistence of partial (local) crystalline order, of long-range solid order, and perfect fluid order. The best agreement is obtained by assuming partial, local order. All results include QP corrections.}
    \label{tab:copper3}
\end{table}

\subsection{Liquid aluminum}\label{subsec43}

The equation curve fitted by Gathers to the enthalpies measured between the melting temperature 933 K and 4000 K is
\begin{equation*}
    H(\mathrm{MJ.kg}^{-1})=\ 4.8910\times 10^{-2}+1.0704\times 10^{-3}~T\nonumber\\
    +2.3084\times 10^{-8}~T^2,
\end{equation*}
from which we derive the following equation for the experimental specific isobaric heat constant $C_P(T)$ :
\begin{equation*}\label{Cp_Gathers_Al}
    C_P(\mathrm{J.g^{-1}.T^{-1}})=1.0704+4.6168\times 10^{-5} T.
\end{equation*}
Table~\ref{tab:aluminum1} displays the mean ion charges $Z^*$ [Eq.~(\ref{Z*_More})], the screening parameter $\kappa$ [Eq.~(\ref{kappa})], and the ion coupling constant $\Gamma$ [Eq.~(\ref{Gamma})] calculated for the temperatures and densities given in the two first columns. 
\begin{table}[!ht]
    \centering
    \begin{tabular}{c c c c c c}
    \hline\hline
     $T$ (K) & $\rho$ (g.cm$^{-3}$) & $Z^*$ & $\kappa$ & $\Gamma$ \\
     \hline
    933 & 2.417 & 2.36 & \hspace{0.15cm}3.17\hspace{0.15cm} & 590 \\
    1000 & 2.402 & 2.32 & 3.17 & 546 \\
    1500 & 2.286 & 2.27 & 3.19 & 343 \\
    2000 & 2.175 & 2.22 & 3.20 & 242 \\
    2500 & 2.075 & 2.17 & 3.21 & 183 \\
    3000 & 1.969 & 2.12 & 3.23 & 143 \\
    3500 & 1.873 & 2.08 & 3.24 & 115 \\
    4000 & 1.775 & 2.03 & 3.25 & 94 \\
    \hline\hline
    \end{tabular}
    \caption{Liquid aluminum: the two first columns report the temperatures and densities of aluminum in the experiments of Gathers \cite{Gathers1983}. The third columns gives the corresponding ion charges $Z^*$ (Eq.~(\ref{Z*_More})), and the fourth and fifth, respectively the corresponding YOCP parameters $\kappa$ and $\Gamma$ [Eqs.~(\ref{kappa}) and (\ref{Gamma})].
    }
    \label{tab:aluminum1}
\end{table}
Table~\ref{tab:aluminum2} presents the values of the Debye temperature $\theta_D$, of $\theta_F$ [using Eqs.~(\ref{Khrapak_Ginf}) and (\ref{Khrapak_eta})], of the Gr\"uneisen constant $\gamma_G$ [Eq.~(\ref{Khrapak_gruneisen})], and of the expansion coefficient $\alpha_V$ [Eq.~(\ref{alpha_V_QP})]. Finally, Table~\ref{tab:aluminum3} shows the results that we obtained for the isobaric heat capacity of liquid aluminum assuming persistence of partial order, long-range order, or the absence of any crystal-type ordering.

\subsubsection{Assessment of the accuracy of $\omega_F$, $\gamma_G$ and $\alpha_V(T)$ obtained using the YOCP model.}

\begin{figure}[!ht]
    \centering
    \includegraphics[width=0.75\linewidth]{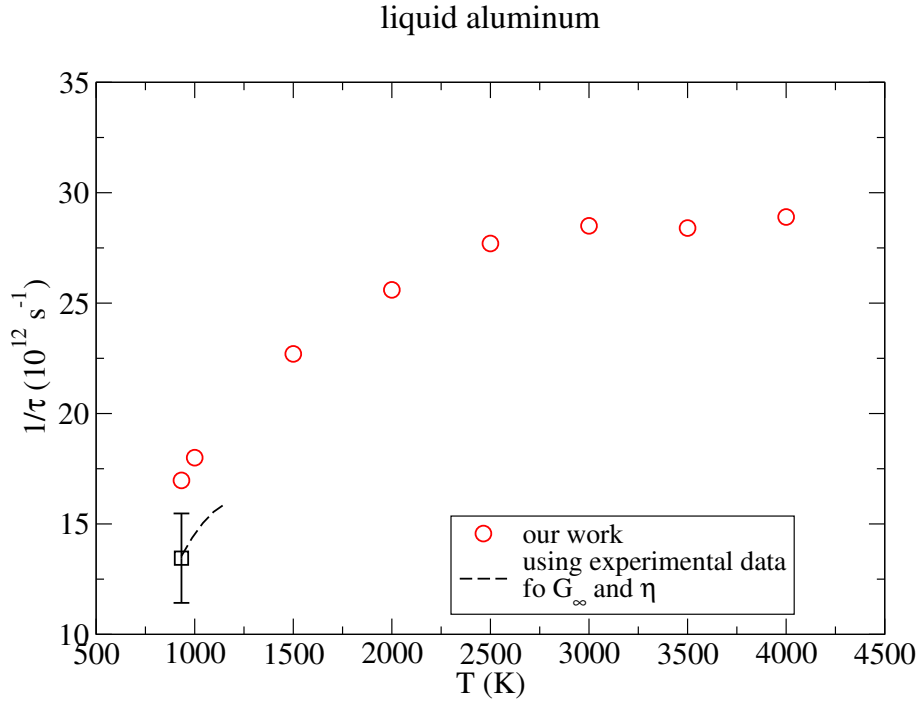}
    \caption{The figure shows the YOCP ratio $G_\infty/\eta$ (red dots) and compares the values at temperatures close to melting to estimations (black square and dashed line) deduced from the available experimental data (Gerlich and Fisher \cite{Gerlich1969}, Tallon and Wolfenden \cite{Tallon1979} and Sutton \cite{Sutton1953} for $G_\infty$ and Assael \emph{et al.}'s fit of experimental shear viscosity between 933 K and 1270 K  \cite{Assael2006}). The error bar on the black square correspond to an estimated 15\% incertitude on the dashed line.
    }
    \label{fig:aluminum5}
\end{figure}

Assael \emph{et al.} fitted the experimental aluminum viscosities within the temperature range 933 K$\leq T\leq$1270 K by the following equation \cite{Assael2006}:
\begin{equation*}
    \log_{10}(\eta/\eta_0)=-a_1+\dfrac{a_2}{T}, 
\end{equation*}
where $a_1=0.7324$, $a_2=803.49$ K and $\eta_0=1$ mPa.s. The standard deviation is 13.7\% within the confidence level of 95\%.  
To the best of our knowledge, the available experimental elastic constants for solid aluminum are those from Gerlich and Fisher \cite{Gerlich1969}, Tallon and Wolfenden \cite{Tallon1979}, and Sutton \cite{Sutton1953}. The three sets of $C_{44}$ data are fitted by the following linear relation, with $C_{44}$ expressed in GPa and $T$ in K:
\begin{equation*}
    C_{44}(T)=32.933-0.015881\,T.
\end{equation*}
The standard error is of 0.35\% on the ``y-intercept'' and of 1.2\% on the slope, within the confidence level of 95\%. According to this, we estimate a probable 2\% error on our estimates of $G_\infty$ obtained by extending this linear relation up to $T\approx$ 1000 K. We add this to the 13.7\% error estimated by Assael \emph{et al.} on the viscosities to obtain an uncertainty estimate for our evaluations of $1/\tau=\omega_F$ in the liquid state from experimental shear moduli and viscosities. The black dashed line in Figure~\ref{fig:aluminum5} represents the ratio of the values of the rhombohedral shear modulus deduced by extrapolation of this fitting curve and the viscosity values given by the approximation formula of Assael \emph{et al.}. With an estimated margin of error of 15-16\%, the YOCP values (depicted as red circles) appear to be approaching the upper limits of the extrapolation error bars, represented with the black square in Figure~\ref{fig:aluminum5}. We also note that, near melting, the variation of $\omega_F$ with temperature is comparable to the estimate in the liquid from extrapolations. 

\begin{table}[!ht]
    \centering
    \begin{tabular}{c c c c c}
    \hline\hline
     $T$ (K) & $\theta_D$ (K) & $\theta_F$ (K) & $\gamma_G$ & $\alpha_V$ (K$^{-1}$) \\
     \hline
    933 & 335 & 129 &\hspace{0.15cm} 1.37 \hspace{0.15cm} & 1.49$\times 10^{-4}$ \\
    1000 & 332 & 136 & 1.35 & 1.44$\times 10^{-4}$ \\
    1500 & 312 & 171 & 1.24 & 1.19$\times 10^{-4}$ \\
    2000 & 294 & 192 & 1.17 & 1.04$\times 10^{-4}$ \\
    2500 & 278 & 205 & 1.11 & 9.41$\times 10^{-5}$ \\
    3000 & 263 & 213 & 1.07 & 8.69$\times 10^{-5}$ \\
    3500 & 249 & 218 & 1.04 & 8.12$\times 10^{-5}$ \\
    4000 & 236 & 222 & 1.01 & 7.76$\times 10^{-5}$ \\
    \hline\hline
    \end{tabular}
    \caption{Liquid aluminum: the table gives the values of the Debye temperature $\theta_D$, of $\theta_F=\hbar\omega_F/k_B$, of the YOCP Gr\"uneisen constant $\gamma_G$ and of the YOCP thermal expansion parameter $\alpha_V$ (with QP corrections) corresponding to the screening and ion coupling parameters of Table \ref{tab:aluminum1}.}
    \label{tab:aluminum2}
\end{table}

\begin{table}[!ht]
    \centering
    \begin{tabular}{c c c c c}\hline\hline
     $T$(K) & $C_P^\mathrm{exp}$ & $C_P^\mathrm{po}$ & $C_P^\mathrm{lro}$ & $C_P^\mathrm{pfo}$ \\
    & \multicolumn{4}{c}{(J/g/K)}\\
     \hline
    933 & \hspace{0.15cm} 1.113 \hspace{0.15cm} & \hspace{0.15cm} 1.165 \hspace{0.15cm} & \hspace{0.15cm} 1.187 \hspace{0.15cm} & \hspace{0.15cm} 0.802 \hspace{0.15cm} \\
    1000 & 1.117 & 1.170 & 1.197 & 0.809\\
    1500 & 1.140 & 1.192 & 1.258 & 0.856\\
    2000 & 1.163 & 1.197 & 1.313 & 0.899\\
    2500 & 1.186 & 1.194 & 1.363 & 0.939\\
    3000 & 1.209 & 1.182 & 1.413 & 0.979\\
    3500 & 1.232 & 1.162 & 1.460 & 1.018\\
    4000 & 1.255 & 1.131 & 1.508 & 1.057\\
    \hline\hline
    \end{tabular}
    \caption{Liquid aluminum: the table compares our $C_P$ calculations (columns 3,4 and 5) to the values derived from the experimental enthalpies of Gathers (column 2). The superscripts "po", "lro", and "pfo" denote the following cases, respectively: persistence of partial (local) crystalline order, of long-range solid order, and perfect fluid order. The best agreement is obtained by assuming partial, local order. All results include QP corrections.}
    \label{tab:aluminum3}
\end{table}

\begin{figure}[!ht]
    \centering
    \includegraphics[width=0.75\linewidth]{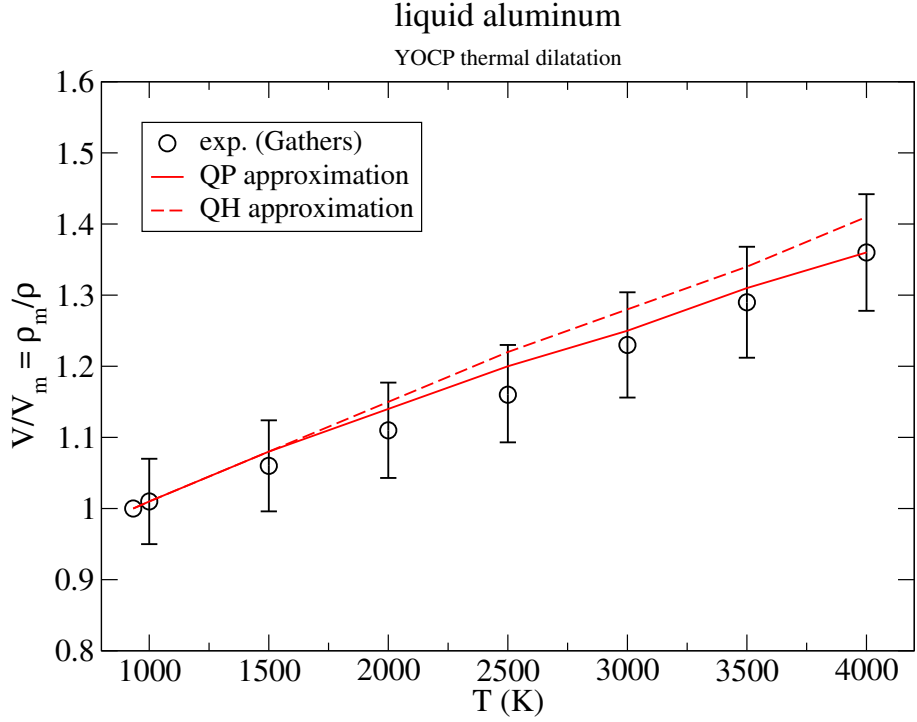}
    \caption{Liquid aluminum volume expansion at constant pressure. Black points: experimental values of Gathers \cite{Gathers1983}. Error bars: 6\%. Red full curve: our work, where we applied QP corrections to the YOCP thermal dilatation coefficient $\alpha_V$. Red dashed curve: the same work, without QP corrections.}
    \label{fig:aluminum2}
\end{figure}

Figure~\ref{fig:aluminum2} compares the volume expansion $V(T)$  derived from the YOCP thermal dilatation constant $\alpha_V$ with the experimental values of Gathers, shown by the black circles with their 6\%  error bars.
The red solid line is obtained using thermal expansion parameters that include QP corrections. The red dashed line is obtained when these corrections are neglected. We observe that the YOCP model can describe thermal expansion and that considering anharmonic effects on the QP scale significantly improves the agreement with experimental values.

Using the Gilev semi-empirical expression given in Eq.~(\ref{Gruneisen_Gilev}) with the parameters $\delta=-2$ and $\beta=10^{-4}$ K$^{-1}$ obtained for liquid aluminum \cite{Gilev2020}, one obtains the estimate $\gamma_G^\mathrm{Gilev}=1.77$ for the value of liquid aluminum's Gr\"uneisen parameter. This is notably higher than the value $\gamma_G=1.37$ obtained using Eq.~(\ref{Khrapak_gruneisen}).

The error  ${\Delta\gamma_G}/{\gamma_G}$ on the Gr\"uneisen parameter will cause an error ${\Delta C_P}/{C_P}$, which we estimate to be given by:
\begin{equation*}
\dfrac{\Delta C_P}{C_P}\approx \dfrac{3\left(\theta_F/\theta_D \right)^3}{3-\left(\theta_F/\theta_D \right)^3}\dfrac{\Delta \theta_D}{\theta_D} + \dfrac{\gamma_G\alpha_V T}{1+(\gamma_G+1/2)\alpha_V T}\dfrac{\Delta\gamma_G}{\gamma_G},   
\end{equation*}
where we have taken the classical limit $C_V^H=Nk_B\left[3-\left(\theta_F/\theta_D \right)^3\right]$ of the harmonic heat capacity at constant volume (justified by the fact that $k_BT\gg\theta_D>\theta_F$), and neglected the impact of ${\Delta\gamma_G}/{\gamma_G}$ on $\alpha_V$ (justified by the fact that the calculated $\alpha_V(T)$ yield thermal volume expansion that agrees with experiments, as shown in Fig.~\ref{fig:aluminum2}). The error in $\gamma_G$ leads to an error in $\theta_D$. Using the definition of the Gr\"uneisen parameter, it can be estimated by the following way:
\begin{equation*}
\gamma_G+\Delta\gamma_G\approx -\dfrac{\partial\ln(\theta_D+\Delta\theta_D)}{\partial\ln V},    
\end{equation*}
\emph{i.e.:}
\begin{equation*}
\Delta\gamma_G\approx -\dfrac{\partial\ln(1+\Delta\theta_D/\theta_D)}{\partial\ln V}.    
\end{equation*}
Then, since $\Delta\theta_D^0=0$, and using once more the definition of $\gamma_G$, one has
\begin{equation*}
1+\dfrac{\Delta\theta_D}{\theta_D} \approx \left(\dfrac{V}{V_0}\right)^{-\Delta\gamma_G} \approx  \left(\dfrac{\theta_D}{\theta_D^0}\right)^{\Delta\gamma_G/\gamma_G}. 
\end{equation*}
Close to the melting temperature, one can further write
\begin{equation*}
\dfrac{\Delta\theta_D}{\theta_D} \approx \left(\dfrac{\theta_D}{\theta_D^0}\right)^{\Delta\gamma_G/\gamma_G}-1 \approx \dfrac{\Delta\gamma_G}{\gamma_G}\times\ln\left(\dfrac{\theta_D}{\theta_D^0} \right),   
\end{equation*}
which finally gives
\begin{equation*}
\dfrac{\Delta C_P}{C_P}\approx \dfrac{\Delta\gamma_G}{\gamma_G} \left[\dfrac{3\left(\theta_F/\theta_D \right)^3}{3-\left(\theta_F/\theta_D \right)^3}\ln\left(\dfrac{\theta_D}{\theta_D^0} \right) \right.\\
\left. +\dfrac{\gamma_G\alpha_V T}{1+(\gamma_G+1/2)\alpha_V T}\right].  
\end{equation*}
We can already see that the two terms on the right-hand side of the equation offset each other to some extent. Using the values reported in Table \ref{tab:aluminum2} for liquid aluminum at 1350 K, we estimate an error $\Delta C_P/C_P\approx 0.15\ \Delta\gamma_G/\gamma_G \approx 5$\% on the value of $C_P(T)$ at the melting temperature. 

\subsubsection{Application to $C_P(T)$}

\begin{figure}[!ht]
    \centering
    \includegraphics[width=0.75\linewidth]{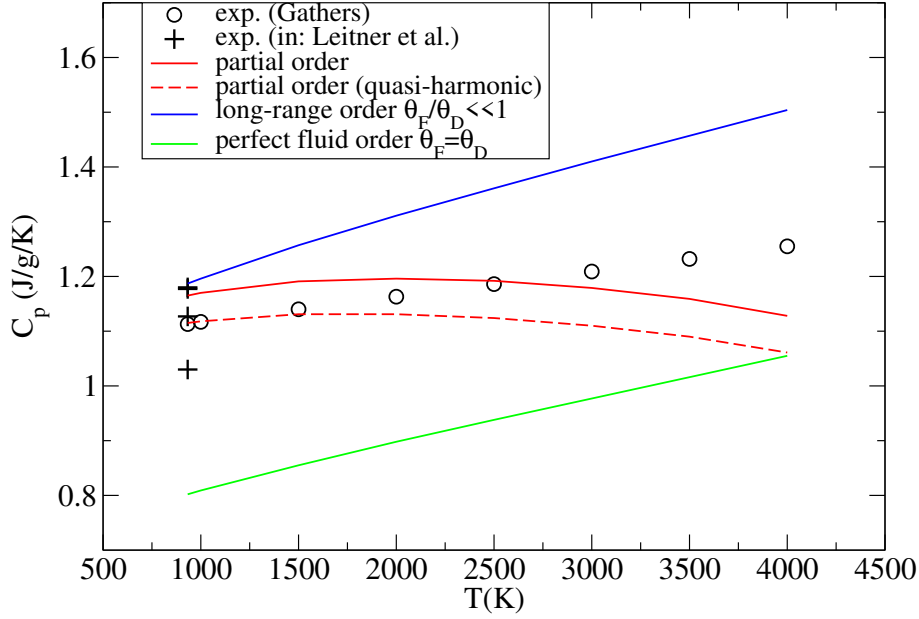}
    \caption{Liquid aluminum heat capacity $C_P$ at constant pressure. Black circles: experimental values of Gathers \cite{Gathers1983}. Black ``+ symbols'': a selection of experimental values by Leitner \emph{et al.} \cite{Leitner2017}. Red curve: our work, assuming partial crystalline order. Blue curve: the same work, considering persistence of long-range crystalline order at any temperature. Green curve: no crystalline order persists in the liquid. All of the solid lines are obtained using the quasi-particle approximation. The red dashed line illustrates the impact of neglecting the QP corrections.}
    \label{fig:aluminum1}
\end{figure}

Figure~\ref{fig:aluminum1} shows the isobaric heat capacities $C_P(T)$ obtained in three configurations (illustrated by the lines) and compares them to the experimental values (circles). As for copper, the green line represents the heat capacity of an ideal, perfectly fluid liquid. The blue line shows the case in which long-range, solid-type ordering persists at all temperatures.  The red curves consider that crystalline ordering only persists partially, gradually disappearing at a rate determined by the ratio $\theta_F/\theta_D$. The solid red line includes QP corrections, while the dashed red one omits them. The values of $C_P(T)$ illustrated by the solid red, blue and green curves are given in Table \ref{tab:aluminum3}.

A relatively flat curve indicates the progressive loss of transverse modes. An ascending curve means that long-range, solid-type order is maintained at high temperatures or that it disappeared as soon as melting began. However, for liquid aluminum, the two hypotheses result in curves that are too high or too low, respectively. Only the red line, which considers the gradual disappearance of local crystalline order, is at the correct height. The fact that this model does not explain the increase in $C_P(T)$  observed experimentally could mean that anharmonic effects are underestimated in our calculation. However, we note that the $T^2$ term in the expression that fits the experimental enthalpies of liquid aluminum is negligible compared to the linear term. At most, it reaches 1.8\% of the linear term's value at 4000 K. Therefore, it appears dispensable in the fitting of the experiments, given the 4\% experimental uncertainty reported by Gathers on the enthalpies. When keeping the $T^2$ term in the fitting, the corresponding coefficient is likely subject to significant uncertainty, which mitigates the discrepancy between the experimental values (black circles) and our results which assume the persistence of local order in the liquid (red curve).

\subsection{First conclusions and outlook}\label{subsec45}

In this first part of our article, we extended the liquid-phonon theory of Bolmatov \emph{et al.} by introducing the YOCP model along with anharmonicity at the QP level. Recent developments in isomorph theory justify the application of the YOCP to dense liquids, expanding its scope beyond the historical framework limited to plasmas and gases. The YOCP is identified as a R-simple system, similar to ones based on Lennard-Jones, Integer-Power-Law  and Exponential type potentials. These systems share scaling laws, which Khrapak \emph{et al.} used to derive the compact, practical expressions that we use in this study. 
Allen's QP approach to anharmonicity restores the thermodynamic consistency that is lost when QH phonon vibrations are replaced by temperature-shifted ones in the free energy rather than in the entropy. 

To evaluate our approach of associating liquid-phonon theory with the YOCP model and anharmonicity at the QP level, we studied the isobaric heat capacities of copper and aluminum. Gathers performed measurements of these materials from melting to 3500 and 4000 K, respectively. Our results demonstrate the relevance of the YOCP-derived phonon relaxation times and thermal expansion parameters in our study. We also illustrate the impact of including anharmonic effects on the heat capacities and the volume temperature expansion. The introduction of anharmonic effects into the formalism has an undeniable impact on the calculation of heat capacities at constant pressure $C_P(T)$ for both copper and aluminum. The QP approach yields excellent agreement with high-temperature experiments for copper. For aluminum, it provides a significant improvement over the QH approximation, though it falls short in fully accounting for the experimental findings. The discrepancies observed between the experimental values and our calculations cannot be attributed to uncertainty in the Gr\"uneisen parameter value, even though this uncertainty is significant (greater than 30\%) for liquid aluminum at melting temperature. We have shown, through an uncertainty analysis, that this has a limited impact on the calculated heat capacities $C_P$. We believe that the apparent increase in experimental $C_P$ with temperature results from the introduction of a quadratic term in the fit of the experimental enthalpies, which is too small compared to the linear term to be considered sufficiently accurate to give a realistic temperature variation of $C_P(T)$, and puts into perspective the hypothesis of very strong impacts of anharmonicity. 

We clearly conclude that liquid-phonon theory best describes the vibrational properties of liquid metals. Anharmonic effects are significant and can be fairly accurately evaluated using the QP approximation. We apply this formalism, which combines liquid-phonon theory with YOCP and anharmonicity at QP level, to calculate the Debye-Waller factors, playing a crucial role in our average-atom approach to compute the electrical resistivity of liquid metals. 

\section{Interpretation of electrical resistivity experiments on liquid copper and liquid aluminum}\label{sec5}

\subsection{Ziman's formula and the Average-Atom model}\label{subsec51}

Although this is not necessary under the thermodynamic conditions considered in this work, the following formulas will be given in the relativistic formalism, for the sake of consistency with the relativistic AA code {\sc Paradisio} \cite{Penicaud2009} that was used to provide the input needed for resistivity calculations. All formulas will be given in atomic units (i.e., $e=\hbar=m_e=1$).
The Ziman formulation of the electrical resistivity \cite{Ziman1961} describes, within the linear response theory, the acceleration of free electrons in a metal and their scattering by an ion. The Ziman resistivity reads then
\begin{equation}\label{eta}
    \rho_\mathrm{dc}^Z=-\dfrac{1}{3\pi {Z^*}^2 n_i} \int_0^\infty \dfrac{\partial f}{\partial \epsilon}(\epsilon,\mu^*) I(\epsilon)\,\mathrm{d}\epsilon,
\end{equation}
where $n_i$ is the ion density, and $Z^*$ the mean ionic charge. The Fermi-Dirac distribution and its derivative read respectively
\begin{equation*}
    f(\epsilon,\mu^*)=\dfrac{1}{e^{\beta (\epsilon-\mu^*)}+1}
\end{equation*}
and
\begin{equation*}
    \dfrac{\partial f}{\partial\epsilon}(\epsilon,\mu^*)=-\beta f(\epsilon,\mu^*)\left[1-f(\epsilon,\mu^*)\right], 
\end{equation*}
where $\beta=1/(k_B T)$ ($k_B$ being the Boltzmann constant) and $\mu^*$ denotes the chemical potential associated to the free electron gas of density $n_e=Z^* n_i$, given by
\begin{equation*}
    \dfrac{2}{(2\pi)^3} \int_0^\infty f(\epsilon,\mu^*)\,4\pi k^2\,\mathrm{d}k=Z^*,
\end{equation*}
or
\begin{equation}\label{fermidirac}
    \mathscr{F}_{1/2}(\beta\mu^*)=\dfrac{\pi^2}{\sqrt{2}}\beta^{3/2}Z^*,
\end{equation}
where 
\begin{equation*}
    \mathscr{F}_{1/2}(x)=\int_0^\infty \dfrac{t^{1/2}}{(1+e^{t-x})}\,\mathrm{d}t.
\end{equation*}
The function $I(\epsilon)$ is given by
\begin{equation*}
    I(\epsilon)=\int_0^{2k}q^3 S(q) \Sigma(q)\,\mathrm{d}q,
\end{equation*}
where $S(q)$ denotes the static ion-ion structure factor and $\Sigma(q)$ the scattering cross-section. The vector $\vec{q}=\vec{k}^\prime-\vec{k}$ is the momentum transferred in the elastic scattering event (i.e. such as $|\vec{k}^\prime|=|\vec{k}|$) of a conduction electron from an initial state $\vec{k}$ to a final $\vec{k}^\prime$ one. Introducing the scattering angle $\theta\equiv (\vec{k},\vec{k}^\prime$) and its cosine $\chi=\cos\theta$, one has $q^2=2k^2 (1-\chi)$ and
\begin{equation*}
    I(\epsilon)=2k^4 \int_{-1}^1 S\left[k\sqrt{2(1-\chi)}\right]|a(k,\chi)|^2 (1-\chi)\,\mathrm{d}\chi.
\end{equation*}
Energy $\epsilon$ and momentum $k$ are related (within the relativistic formalism, $c$ being the speed of light, and using atomic units) by
\begin{equation*}
    k=\sqrt{2\epsilon \left(1+\dfrac{\epsilon}{2c^2}\right)}.
\end{equation*}
The $t-$matrix formalism of Evans \cite{Evans1973} provides the electron-ion scattering amplitude $|a(k,\chi)|$, whose square is actually $\Sigma(q)$, given by, in the relativistic framework \cite{Sterne2007}
\begin{equation}\label{scattering}
    |a(k,\chi)|^2=\frac{1}{k^2}\left(\Big|\sum |\kappa|e^{i\delta_\kappa(k)}\sin[\delta_\kappa(k)]P_{\ell}(\chi)\Big|^2\right.\nonumber\\
    +\left.\Big|\sum \frac{|\kappa|}{i\kappa}e^{i\delta_\kappa(k)}\sin[\delta_\kappa(k)]P^1_\ell(\chi)\Big|^2\right),
\end{equation}
where summations are performed on the electronic states, labeled by the relativistic quantum number $\kappa$, which is related to the quantum number $\ell$ associated with the orbital momentum $L$ and with the spin $s$ by the relations
\begin{equation*}
    \begin{array}{l c l}
        \kappa=-(\ell+1)& \mathrm{ for }& s=+1/2,\\
        \kappa=\ell& \mathrm{ for }& s=-1/2.
    \end{array}
\end{equation*}
The functions $P_\ell$ and $P^1_\ell$ denote respectively the Legendre and associated Legendre polynomials.

The average ion charge $Z^*$ and phase-shifts $\delta_{\kappa}(k)$ needed in Eqs.~(\ref{eta}), (\ref{fermidirac}) and (\ref{scattering}), respectively, can be obtained with the help of AA codes. The ionic structure factor $S(k)$ is usually obtained independently. Like in our preceding studies, we apply in the present work the method developed by Rogers to solve the hyper-netted chain (HNC) equation for charged spheres \cite{Rogers1980}. The semi-empirical correlation function $g(r)$ from Held and Pignolet \cite{Held1986} is used to initialize the iteration process.

In Ref.~\cite{Wetta2020} we show that this HNC structure factor allows a continuous description of electrical resistivity from the solid to the plasma states when an additional contribution $\delta\rho_\mathrm{dc}$ is added to the Ziman resistivity [Eq.~(\ref{eta})], given by:
\begin{equation}\label{correction_resistivité}
    \delta\rho_\mathrm{dc}=-\dfrac{1}{3\pi {Z^*}^2 n_i}\sum_G \dfrac{N(G)}{4\pi}\mathrm{e}^{-2WG^2}\\
    \int_{G/2}^\infty \left(-\dfrac{\partial f}{\partial k} \right) k^2 G^2\times\Bigg|a\left(k,1-\dfrac{G^2}{2k^2}\right)\Bigg|^2\,\mathrm{d}k.\nonumber
\end{equation}
$N(G)$ denotes the number of reciprocal lattice vectors of same length $G$ and $\mathrm{e}^{-2WG^2}$ the Debye-Waller factors accounting for thermal decay of crystalline-type order. This correction results from the extension to liquids of a prescription by Rosenfeld and Stott initially formulated for solids \cite{Rosenfeld1990}, which consists in removing the contribution of the perfect rigid lattice from the total structure factor used in Ziman's formula. As mentioned in the introduction, this concept has been previously applied by Baiko {\it et al.} in the framework of astrophysics \cite{Baiko1998}. An experimental proof of transient long-range order in melted gold has been presented \cite{Mo2018} through the coexistence of Debye-Scherrer rings and Laue diffraction peaks in x-ray diffraction patterns at times exceeding the electron-ion equilibration one. 

In this section, we aim to bridge the gap between the expressions for $2W$ established specifically for solids and plasmas, providing a global expression applicable regardless of how the Ziman resistivity is obtained (using AA codes or other methods). The formalism developed and applied to $C_P$ fulfills both constraints.
\\

\subsection{The Debye-Waller factor in the framework of the liquid phonon theory}\label{subsec52}

\subsubsection{Harmonic Debye-Waller factor in the Debye model}

In the harmonic approximation, the Debye-Waller factor $e^{-2WG^2}$ is related to the mean squared atomic displacement. The argument $2WG^2$ of the exponential function reads
\begin{equation*}
    2WG^2=\left\langle \left(\vec{G}\cdot\vec{u_l} \right)^2 \right\rangle,
\end{equation*}
where $\vec{u_l}$ represents the displacement from its equilibrium position of an atom on the lattice site $l$. The angular brackets denote the thermal average. As in the previous section, $\vec{G}$ is a vector of the reciprocal lattice. $2WG^2$ then reads \cite{Kittel1965}
\begin{align}\label{DWaller}
    2WG^2 &=\dfrac{\hbar}{2m_i N} \sum_{\vec{k},j} \dfrac{\left[\vec{G}\cdot\vec{e}_{\vec{k},j}\right]^2}{\omega_{\vec{k},j}} \left[1+2n(\omega_{\vec{k},j})\right]\nonumber\\
          &=\dfrac{\hbar}{2m_i N} \sum_{\vec{k},j} \dfrac{\left[\vec{G}\cdot\vec{e}_{\vec{k},j}\right]^2}{\omega_{\vec{k},j}}\coth\left(\dfrac{\beta\hbar\omega_{\vec{k},j}}{2}\right).
\end{align}
The sums are performed on harmonic phonon frequencies $\omega_{\vec{k},j}$, where $\vec{k}$ denote wave vectors, $j$ polarization indices, and where $\vec{e}_{\vec{k},j}$ are eigenvectors.  For cubic systems, one finds
\begin{equation*}
    2WG^2=\dfrac{1}{3}\langle u_l^2\rangle G^2=\dfrac{\hbar G^2}{6m_i N} \sum_{\vec{k},j} \dfrac{\coth\left(\dfrac{\beta\hbar\omega_{\vec{k},j}}{2}\right)}{\omega_{\vec{k},j}}, 
\end{equation*}
and thus
\begin{equation*}
    2W=\dfrac{\hbar}{6m_i N} \sum_{\vec{k},j} \dfrac{\coth\left(\dfrac{\beta\hbar\omega_{\vec{k},j}}{2}\right)}{\omega_{\vec{k},j}}. 
\end{equation*}
Throughout the rest of our article, we will adopt the common practice of referring to the quantity $2W$ as the Debye-Waller factor. In the framework of the liquid-phonon theory, the harmonic Debye-Waller factor reads:
%
%\begin{widetext}
\begin{align*}
    2W^H &=\dfrac{\hbar}{6m_i N} \left\{\int_0^{\omega_D} \dfrac{\coth\left(\dfrac{\beta\hbar\omega}{2}\right)}{\omega}\dfrac{3N\omega^2}{\omega_D^3}\,\mathrm{d}\omega + \int_{\omega_F}^{\omega_D} \dfrac{\coth\left(\dfrac{\beta\hbar\omega}{2}\right)}{\omega}\dfrac{6N\omega^2}{\omega_D^3}\,\mathrm{d}\omega \right\} \\
            &=\dfrac{\hbar^2 G^2}{m_i k_B\theta_D} \left\{ \dfrac{1}{4}\left(3-2\left(\dfrac{\theta_F}{\theta_D}\right)^2\right) + \dfrac{T}{\theta_D}\left[3D_1\left(\dfrac{\theta_D}{T}\right)-2\left(\dfrac{\theta_F}{\theta_D}\right)D_1\left(\dfrac{\theta_F}{T}\right)\right] \right\},
\end{align*}
%\end{widetext}
%
where $D_1(x)$ is the Debye function of order 1, defined by the following equation:
\begin{align*}
    D_1(x) &=\dfrac{1}{x} \int_0^x \dfrac{x}{e^x-1}\,\mathrm{d}x\\
           &=\dfrac{1}{x}\left[\dfrac{\pi^2}{6}+x \ln(1-e^{-x})-\mathrm{Li}_2(e^{-x})\right].
\end{align*}
In the classical limit where $T\gg\theta_D$ and $\theta_F$, the expression takes the simple form:
\begin{equation*}
    \lim_{T\gg\theta_D\geq\theta_F} 2W^H= \dfrac{\hbar^2}{m_i k_B\theta_D} \dfrac{T}{\theta_D} \left(3-2\dfrac{\theta_F}{\theta_D}\right).
\end{equation*}

\subsubsection{Anharmonic corrections in the quasi-particle approximation}

According to Pathak and Deo \cite{Pathak1967}, the anharmonic Debye-Waller factor is obtained by replacing the harmonic phonon frequency $\omega_{\vec{k},j}$ by the shifted frequency $\omega_{\vec{k},j}+\Delta\omega_{\vec{k},j}$ in Eq.~(\ref{DWaller}). In the quasi-particle approximation, this is the frequency denoted by $\omega(V,T)$ in Sec.~\ref{subsec24}, verifying the following relation:
\begin{equation*}
    \left.\dfrac{\partial\ln\omega(V,T)}{\partial T}\right|_P = -\left(\dfrac{1}{2}+\gamma_G\right) \alpha_V(T).
\end{equation*}
The shifted frequency is then given by
\begin{equation*}
    \omega(V,T)=\omega(V[T]) \exp\left[-\int_{T_0}^{T} \left(\dfrac{1}{2}+\gamma_G\right) \alpha_V(T)\,\mathrm{d}T \right].
\end{equation*}
In this formula, we set $T_0 = T_m$. This simplification assumes that the implicit temperature dependence of the quasi-harmonic frequencies dominates the explicit temperature dependence accounted for by the QP approximation at melting. Introducing the notation
\begin{equation*}
    f_\omega= \exp\left[-\int_{T_0}^{T} \left(\dfrac{1}{2}+\gamma_G\right) \alpha_V(T)\,\mathrm{d}T \right], 
\end{equation*}
the QP Debye-Waller factor reads

%\begin{widetext}
\begin{equation}\label{DW_QP}
    2W^{QP}= \dfrac{\hbar^2}{m_ik_B\theta_D}\dfrac{1}{f_\omega^2}\left\{\dfrac{1}{4}\left(3-2\left(\dfrac{\theta_F}{\theta_D}\right)^2\right)f_\omega+\dfrac{T}{\theta_D}\left[3D_1\left(\dfrac{\theta_D}{T}f_\omega\right)-2\left(\dfrac{\theta_F}{\theta_D}\right)D_1\left(\dfrac{\theta_F}{T}f_\omega\right)\right] \right\},   
\end{equation}

which reduces, at the limit $T\gg\theta_D\geq\theta_F$, to

\begin{equation*}
    \lim_{T\gg\theta_D\geq\theta_F}  2W^{QP}= \dfrac{\hbar^2}{m_i k_B\theta_D} \dfrac{T}{\theta_D} \left(3-2\dfrac{\theta_F}{\theta_D}\right)\exp\left[2\int_{T_m}^{T} \left(\dfrac{1}{2}+\gamma_G\right) \alpha_V(T)\,\mathrm{d}T \right].
\end{equation*}
%\end{widetext}
%

\subsubsection{Comparison to other models}

For cubic lattices, the DW factors are isotropic and can be calculated  from the phonon density of states (PDOS) $g(\omega)$ for all temperatures. Gao and Peng performed parametrization for the temperature dependence of the Debye-Waller factors for 68 elemental crystals \cite{Gao1999}, integrating experimental PDOS or using the Debye model $g(\omega)=3\omega^2/\omega_D^3$ when the experimental PDOS was unavailable. They have found that the resulting Debye-Waller factors vary smoothly with temperature and accurately fitted them using a fourth-degree regression method, giving the following parameterizations:
\begin{equation}\label{DW_Gao_Peng}
    B=2W\times 8\pi^2=a_0+a_1T+a_2T^2+a_3T^3+a_4T^4.
\end{equation}
Two parameterization series are proposed: one for $T<80$ K and one for temperatures between 80 K and $\min(1000\,\mathrm{K},T_m)$. We use the latter for our comparisons. We extend the ranges of these parameters beyond their valid limits, assuming that the values obtained slightly above the melting temperature remain realistic.

%\begin{widetext}

\begin{figure}[!ht]
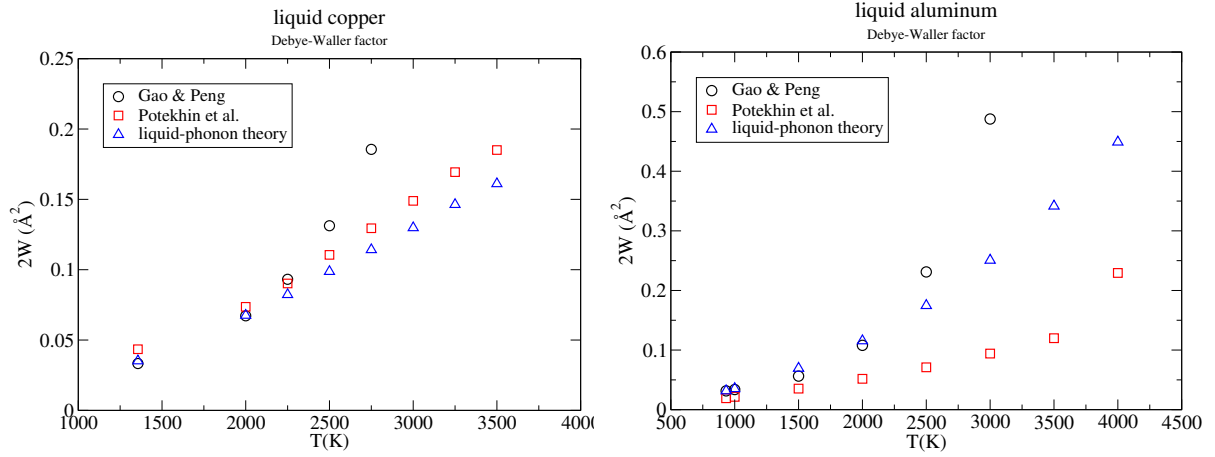

\centering
\begin{subfigure}{.49\textwidth}
  \centering
  \includegraphics[width=\linewidth]{DWaller_Cu.eps}
  \label{fig:sub1}
\end{subfigure}
\begin{subfigure}{.49\textwidth}
  \centering
  \includegraphics[width=\linewidth]{DWaller_Al.eps}
  \label{fig:sub2}
\end{subfigure}
\caption{The figure compares, for liquid copper and liquid aluminum, the Debye-Waller factor $2W$ obtained with liquid-phonon theory [blue triangles, Eq.~(\ref{DW_QP})], with the interpolation formula of Potekhin \emph{et al.} [red squares, Eq.~(\ref{DW_Potekhin})], and extending the Gao and Peng parameterization in the liquid state [black circles, Eq.~(\ref{DW_Gao_Peng})].}
\label{fig:DWaller}
\end{figure}

%\end{widetext}

The DW factor model proposed by Potekhin \emph{et al.} \cite{Potekhin1999} is more suitable for high-temperature comparisons. For cubic crystals, $2 Wq^2=\frac{1}{3} q^2 r_0^2$, where $r_0^2$ is the root-mean square deviation of an ion at a crystal site, given by
\begin{equation*}
    r_0^2=\dfrac{3\hbar^2}{2m_i k_BT}\Big\langle \dfrac{1}{z}\coth\left(\dfrac{z}{2}\right)\Big\rangle,
\end{equation*}
where the authors introduced $z=\hbar \omega/k_BT$, and the angular brackets denote average over phonon frequencies within the Brillouin zone (BZ)
\begin{equation*}
    \Big\langle \dfrac{1}{z}\coth\left(\dfrac{z}{2}\right)\Big\rangle=\dfrac{1}{3V_\mathrm{BZ}}\sum_s \int_\mathrm{BZ} \dfrac{1}{z}\coth\left(\dfrac{z}{2}\right)\,\mathrm{d}\vec{k}.
\end{equation*}
The latter average is related to the frequency moments of the phonon spectrum $u_n=\Big\langle (\omega/\omega_p)^n\Big\rangle$, $\omega_p$ denoting the plasma frequency within the context of fully ionized plasma considered by Potekhin \emph{et al.}
\begin{equation*}
    \omega_p=\left(\dfrac{e^2}{\epsilon_0}\dfrac{{Z}^2n_i}{ m_i}\right)^{1/2}.
\end{equation*}
Introducing the plasma temperature $T_p=\hbar\omega_p/k_B$, the high temperature  limit of the average over phonon frequencies reads
\begin{equation*}
    \lim_{T\gg T_p} \Big\langle \dfrac{1}{z}\coth\left(\dfrac{z}{2}\right)\Big\rangle=2 u_{-2}\times \left(\dfrac{T}{T_p}\right)^2,
\end{equation*}
and the low temperature $T\ll T_p$ asymptote is given by
\begin{equation*}
    \lim_{T\ll T_p} \Big\langle \dfrac{1}{z}\coth\left(\dfrac{z}{2}\right)\Big\rangle= u_{-1}\times \left(\dfrac{T}{T_p}\right).
\end{equation*}
Potekhin \emph{et al.} calculated the bcc (body-centered cubic) Debye-Waller factors for the temperatures ranging from $0.02\,T_p$ up to $20\,T_p$, and electron relativistic parameters $x=\hbar k_F/(m_e c)$ between 0.1 and 30, corresponding to densities $10^3$ g/cm$^3\lesssim \rho \lesssim 10^{11}$ g/cm$^3$ and complete ionization, and fitted an interpolation formula between the two limits $T\ll T_p$ and $T\gg T_p$ \cite{Potekhin1999}:
\begin{equation}\label{DW_Potekhin}
    2 W=\dfrac{\hbar^2}{2 m_i (k_BT_p)}\left[u_{-1}e^{-9.1\,t}+2u_{-2}\,t \right],
\end{equation}
where $t=T/T_p$. For bcc crystals: $u_{-1}=2.8$, and $u_{-2}=13$. However, the authors expect their results to also be applicable at low densities of $1$ g/cm$^3\lesssim \rho \lesssim 10^{3}$ g/cm$^3$ where ionization is incomplete. To that end, they recommend replacing the atomic number $Z$ with the ion charge $Z^*$ in plasma frequency expression $\omega_p$.

The interpolation function reproduces the expected temperature dependencies for solids and liquids. For $t\ll 1$, the exponential term $\exp(-9.1\,t)$ can be expanded in powers of $t$ to give a result similar to the Gao and Peng parameterization. The difference is that the parameters $a_i$ depend on the plasma temperature rather than the Debye temperature. For liquids, the exponential term becomes negligible, and the formula tends to $2W\propto T/T_p^2$. Here also, the plasma temperature replaces the Debye temperature in the expression of $2W$ given by liquid-phonon theory. The interpolation formula developed by Potekhin \emph{et al.} ensures continuity from the plasma to the liquid and solid states. However, this is achieved at the cost of losing connection to atomic vibrations.
The liquid-phonon approach restores this link and is therefore expected to be more relevant in the liquid domain. But its scope of application cannot extend beyond the vaporization curve.

Figure~\ref{fig:DWaller} compares the Debye-Waller factor $2W$ obtained with liquid-phonon theory (represented by blue triangles) for liquid copper and liquid aluminum, with the interpolation formula of Potekhin \emph{et al.} (red squares) and the extended Gao and Peng parameterization in the liquid state (black circles). The Gao and Peng parameterizations are based on Debye-Waller factors obtained by directly integrating experimental PDOS measured at 300 K. As anticipated, the Gao and Peng model diverges rapidly when extended to the liquid state. In liquid-phonon theory, $2W$ grows more slowly than in the Potekhin \emph{et al.} model. Thus, the Bragg peaks are more attenuated, and higher electrical conductivities are expected in the liquid state when liquid-phonon theory is applied. According to the figure, the impact on resistivity calculations is expected to be greater for aluminum than for copper.

\subsection{Liquid aluminum resistivity}\label{subsec53}

At the pressure of 0.3 GPa used of the experiments of Gathers, aluminum crystallizes in a face-centered cubic (fcc) structure. The natural assumption for the crystal order persisting in liquid aluminum is then the fcc-type order. In a previous study of liquid aluminum \cite{Wetta2020} we found that the experimental resistivities are framed by the values calculated with the assumption of partial fcc order and that of partial bcc order. In the same study, we suggested that the locally persistent order could be a tetragonal centered structure with a $c/a$ ratio of the axes between 1 (bcc case) and $\sqrt{2}$ (fcc case), or equivalently, a face-centered tetragonal (fct) structure with a $c/a$ ratio between 1 (fcc) and $1/\sqrt{2}$ (bcc). The present section will examine this assertion.

Using an AA model to evaluate the mean ionic charge can be complicated when the electronic density of states differs significantly from the $n(\epsilon)\propto \sqrt{\epsilon}$ type characterizing perfectly free electron gases. A correct estimate of $Z^*$ is a prerequisite before undertaking the study of the ionic structure of the liquid that follows. This problem does not arise for liquid aluminum at temperatures ranging from the melting point to approximately 3250 K, at which point the AA model converges to solutions with three electrons (``s'' and ``p'') in the energy continuum. We use $Z^* = 3$ for 933 K $\leq T \leq $ 3250 K. Above 3250 K, the AA code {\sc Paradisio} fails to converge to a stable solution. Instead, it oscillates between two unstable configurations with three and two free electrons, respectively. Neither of these two configurations meets the criteria that we set out in \cite{Wetta2025} to assess the relevance of the AA $Z^*$. Consequently, the resistivity at $T = 4000$ K, as measured by Gathers, is not calculated.

The Debye-Waller factors from the liquid-phonon theory [Eq.~(\ref{DW_QP})] are used to calculate the $\delta\rho_\mathrm{dc}$ correction which accounts for the persistence of local order, assumed to be of the fct-type. The $c/a$ axis ratio is adjusted to best align with Gathers's experimental values. We compared our results with the resistivities obtained by the author using the resistance ratio method with a fixed room-temperature geometry for each shot. The thermal expansion effects were reintroduced by multiplying by the ratio of the measured volumes. The estimated error on resistivities is 4\%. 

\begin{figure}[!ht]
    \centering
    \vspace{0.5cm}
    \includegraphics[width=0.75\linewidth]{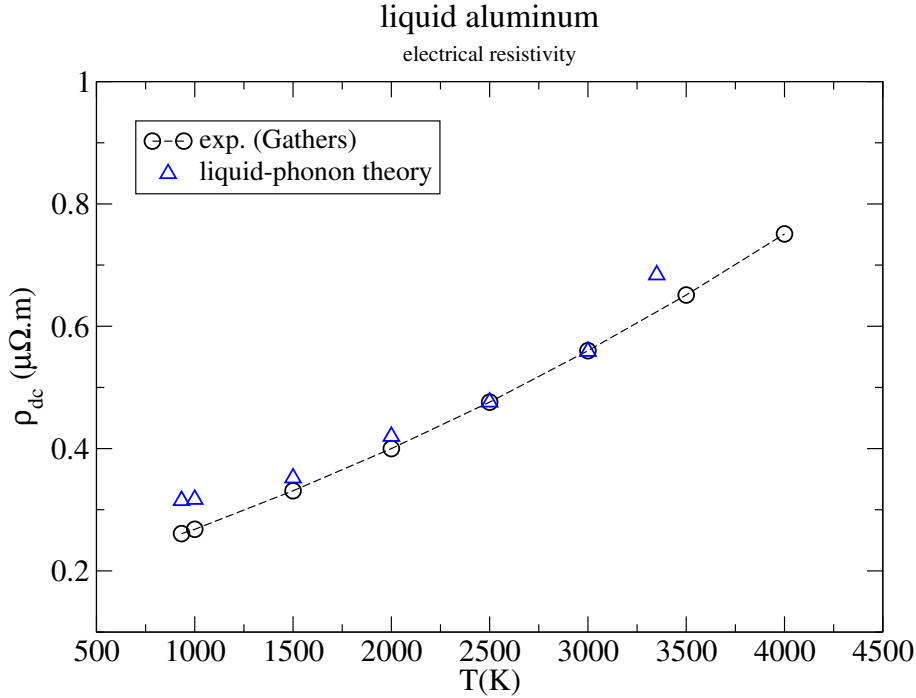}
    \caption{Liquid aluminum electrical resistivity, accounting for persistence of partial fct (face-centered tetragonal) order. Blue triangles: resistivities obtained using the Debye-Waller factors from the liquid-phonon theory [Eq.~(\ref{DW_QP})]. Black circles: experiments \cite{Gathers1983}.}
    \label{fig:Al_resistivity1}
\end{figure}
\begin{figure}[!ht]
    \centering
    \vspace{0.5cm}
    \includegraphics[width=0.75\linewidth]{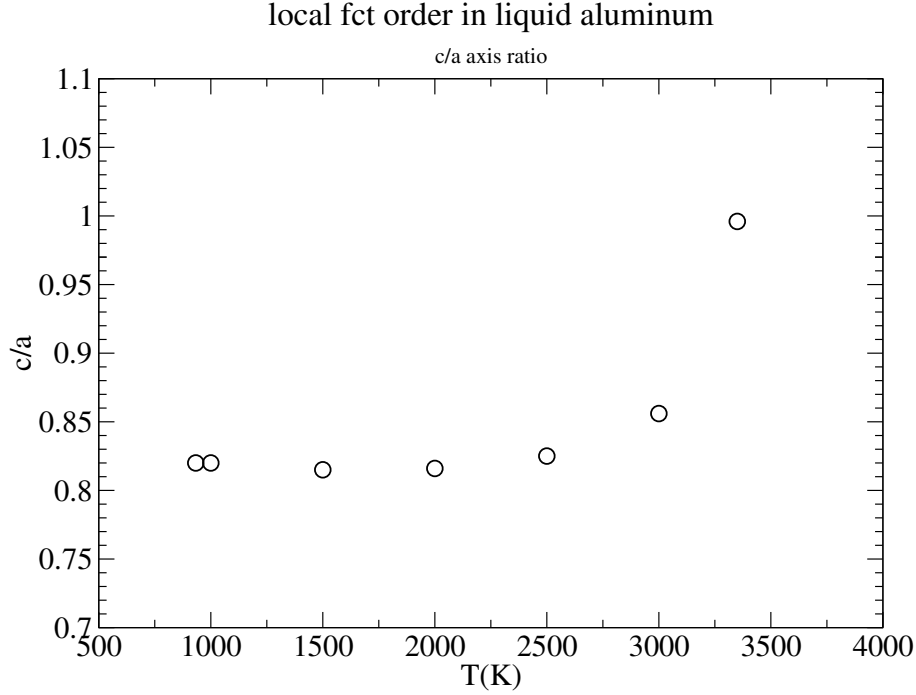}
    \caption{The figure presents the values of the axis length ratio $c/a$ giving the best agreement between our liquid-phonon calculations and experimental values.}
    \label{fig:Al_resistivity2}
\end{figure}

Figure~\ref{fig:Al_resistivity1} shows the electrical resistivity calculated for liquid aluminum, considering the persistence of a local order of the fct-type (blue triangles) and compares them to the experimental values \cite{Gathers1983}, represented by the black circles. Figure~\ref{fig:Al_resistivity2} presents the values of the axis length ratio $c/a$ giving the best agreement between our liquid-phonon calculations and the experimental values. We find it to vary from $c/a\approx 0.8$ for the dense liquid up to $c/a\approx 1$ for the less dense one. 

As mentioned above, the natural assumption for the crystal order that persists in liquid aluminum is the fcc-type order. By this we mean a layered arrangement of the neighboring atoms comparable to that observed in an fcc crystal, \emph{i.e.}, $N=12$ atoms for the nearest neighbors, $N=8$ for the second neighbors, and so on. However, the radii of these atomic layers could fluctuate around the ideal radii of the fcc structure. This is precisely what happens in the fct deformation of the ideal fcc structure. Tab.~\ref{tab:aluminum4} shows how compressing an fcc cell by $r=0.8$ along its vertical axis changes its local environment. The compression changes the crystal structure from fcc to fct, with an axis ratio of $c/a = r$. For $c/a=0.8$, the first four layers of neighboring atoms are slightly split, with the average distances from the central atom, however, remaining close to those of the fcc structure. We interpret these deformations by the need to disrupt the ideal local order to facilitate the flow of atoms from highly ordered areas to liquid ones, in accordance with Frenkel's vision of liquid. This is particularly crucial when the liquid is still dense. As the density decreases, atomic displacements could gradually occur without these small deformations relative to the fcc crystal structure, which would explain why it is no longer necessary to compress the fcc lattice to reproduce the experimental resistivities.

%
%\begin{widetext}

\begin{table}[!ht]
\centering
\begin{tabular}{l  l  l}
\hline\hline
\multicolumn{1}{c}{fcc}& \multicolumn{1}{c}{\begin{tabular}[c]{@{}l@{}}\hspace{1.5cm} fct \\ ($r=\dfrac{c_\mathrm{fct}}{a_\mathrm{fct}}$ and $a_\mathrm{fct}=\dfrac{a_\mathrm{fcc}}{r^{1/3}}$)\end{tabular}} & \multicolumn{1}{c}{fct ($r=0.8$)} \\ \hline
\multicolumn{1}{c}{$N=12$ \hspace{0.5cm} $d=\dfrac{a_\mathrm{fcc}}{\sqrt{2}}\approx 0.707\,a_\mathrm{fcc}$} & \begin{tabular}[c]{@{}l@{}}  $\qquad N=8$ \hspace{0.5cm} $d=\dfrac{a_\mathrm{fct}}{2}\sqrt{1+r^2}\qquad $ \\ \\ $\qquad N=4$ \hspace{0.5cm} $d=\dfrac{a_\mathrm{fct}}{\sqrt{2}}\qquad$ \end{tabular} & \begin{tabular}[c]{@{}l@{}}$\qquad N=8$ \hspace{0.5cm} $d\approx 0.689\,a_\mathrm{fcc}\qquad$ \\ \\ $\qquad N=4$ \hspace{0.5cm} $d\approx 0.761\,a_\mathrm{fcc}\qquad$ \end{tabular}\\ \hline
\multicolumn{1}{c}{$N=6$ \hspace{0.5cm} $d=a_\mathrm{fcc}$} & \begin{tabular}[c]{@{}l@{}}$\qquad N=4$ \hspace{0.5cm} $d=a_\mathrm{fct}\qquad$ \\ \\ $\qquad N=2$ \hspace{0.5cm} $d=c_\mathrm{fct}=r\, a_\mathrm{fct}\qquad$ \end{tabular} & \begin{tabular}[c]{@{}l@{}}$\qquad N=4$ \hspace{0.5cm} $d\approx 1.077\,a_\mathrm{fcc}\qquad$ \\ \\ $\qquad N=2$ \hspace{0.5cm} $d\approx 0.861\,a_\mathrm{fcc}\qquad$ \end{tabular}\\ \hline
\multicolumn{1}{c}{$N=24$ \hspace{0.5cm} $d=a_\mathrm{fcc}\sqrt{\dfrac{3}{2}}\approx 1.225\,a_\mathrm{fcc}$} & \begin{tabular}[c]{@{}l@{}}$\qquad N=16$ \hspace{0.3cm} $d=\dfrac{a_\mathrm{fct}}{2}\sqrt{5+r^2}$ \\ \\ $\qquad N=8$ \hspace{0.5cm} $d=a_\mathrm{fct}\sqrt{\dfrac{1}{2}+r^2}$ \end{tabular} & \begin{tabular}[c]{@{}l@{}}$\qquad N=16$ \hspace{0.3cm} $d\approx 1.279\,a_\mathrm{fcc}$ \\ \\ $\qquad N=8$ \hspace{0.5cm} $d\approx 1.150\,a_\mathrm{fcc}$ 
\end{tabular}\\ \hline
\multicolumn{1}{c}{$N=12$ \hspace{0.5cm} $d=a_\mathrm{fcc}\sqrt{2}\approx 1.414\,a_\mathrm{fcc}$} & \begin{tabular}[c]{@{}l@{}} $\qquad N=8$ \hspace{0.5cm} $d=a_\mathrm{fct}\sqrt{1+r^2}$ \\ \\ $\qquad N=4$ \hspace{0.5cm} $d=a_\mathrm{fct}\sqrt{2}$ \end{tabular} & \begin{tabular}[c]{@{}l@{}}$\qquad N=8$ \hspace{0.5cm} $d\approx 1.396\,a_\mathrm{fcc}$ \\ \\ $\qquad N=4$ \hspace{0.5cm} $d\approx 1.523\,a_\mathrm{fcc}$ \end{tabular}\\ \hline\hline
\end{tabular}
\caption{The table shows how compressing an fcc (face-centered cubic) cell by $r=0.8$ along its vertical axis changes its local environment. This changes the crystal structure from fcc to fct (face-centered tetragonal), with an axis ratio of $c/a = r$.}\label{tab:aluminum4}
\end{table}

%\end{widetext}

\begin{figure}[!ht]
    \centering
    \includegraphics[width=0.75\linewidth]{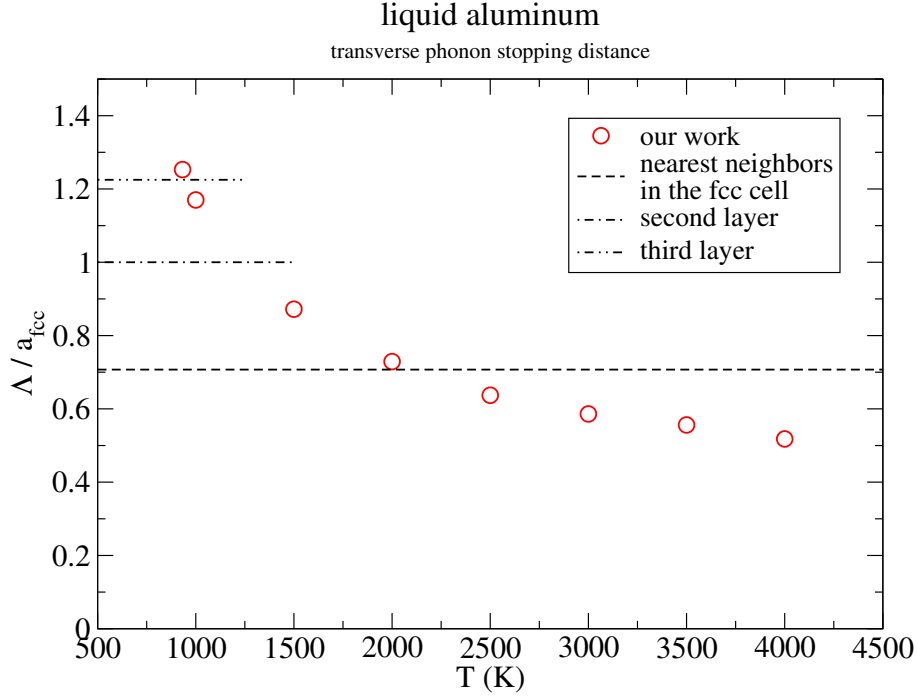}
    \caption{Extent of the partial order in liquid aluminum: the circles represent the typical distances $\Lambda$ at which the amplitude of acoustic shear waves are reduced by 95\%. They are compared to the positions of the three first neighboring shells in a fcc cell.}
    \label{fig:Al_partial_order}
\end{figure}

To clarify the meaning of  ``partial-order'' in liquids, figure \ref{fig:Al_partial_order} shows, for the case of liquid aluminum at the temperatures considered in this work, the distances $\Lambda$ at which shear acoustic waves travel before losing 95\% of their initial amplitude (red circles). They are compared to the positions of the three first neighboring shells (dashed lines), for which we have considered the fcc crystalline order for simplicity's sake. Shear waves being damped by a factor $e^{-t/\tau}$ (where $\tau$ is the Frenkel relaxation time), their amplitude is reduced by a factor 0.95 at $t\approx 3\tau$. The distances $\Lambda$ shown in the figure are given by $\Lambda=3\tau\,v_s$, where $v_s$ denotes the shear wave velocity. Within the Debye model :
\begin{equation*}
v_s=\dfrac{(6\pi^2)^{1/3}}{a_\mathrm{fcc}}\omega_D,    
\end{equation*}
where $\omega_D$ is the Debye frequency
\begin{equation*}
\omega_D=\dfrac{k_B\theta_D}{\hbar},    
\end{equation*}
and $a_\mathrm{fcc}$ is the fcc lattice parameter, given by
\begin{equation*}
a_\mathrm{fcc}=\left(4\times\dfrac{4\pi}{3}R_\mathrm{WS}^3 \right)^{1/3}.    
\end{equation*}

\subsection{Liquid copper resistivity}\label{subsec54}

In Ref.~\cite{Wetta2025}, we addressed the task of evaluating the average ionic charge for a ``3d-block'' metal using an average-atom model. A detailed analysis of the role of different charge densities in electrical conductivity of liquid copper yielded the estimates of $Z^*$ reported in Table \ref{tab:Z*_Cu}, used in our calculations. 

Like aluminum, copper crystallises in the face-centred cubic structure. Our analysis of the resistivity of liquid aluminum reveals that the local fcc-type order undergoes tetragonal deformation beyond melting. We examined the possibility that the same might be true for copper. As for aluminum, we considered a local fcc-type order and adjusted the $c/a$ ratio of the axis to the electrical resistivities measured by Gathers for copper. For this purpose, we used the data obtained by the electrical resistance ratio method. Gathers's estimation of the error bar is 4\%.

\begin{figure}[!ht]
    \centering
    \includegraphics[width=0.75\linewidth]{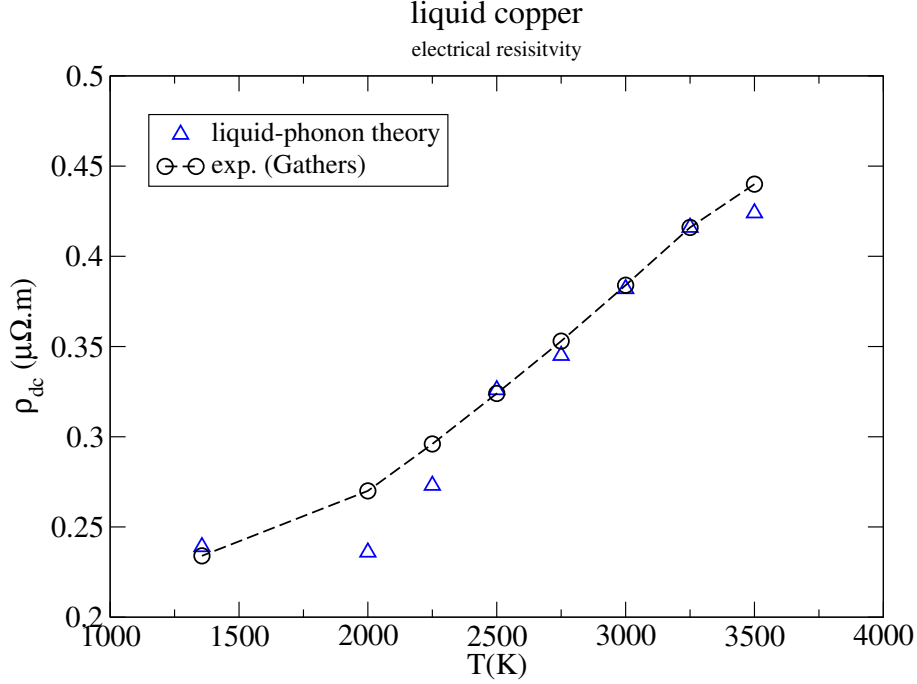}
    \caption{Liquid copper electrical resistivity, accounting for persistence of partial fcc order.  Blue triangles: with Debye-Waller factors from the liquid-phonon theory 
     [Eq.~(\ref{DW_QP})]. Black circles: experiments \cite{Gathers1983}.}
    \label{fig:Cu_resistivity1}
\end{figure}

\begin{figure}[!ht]
    \centering
    \includegraphics[width=0.75\linewidth]{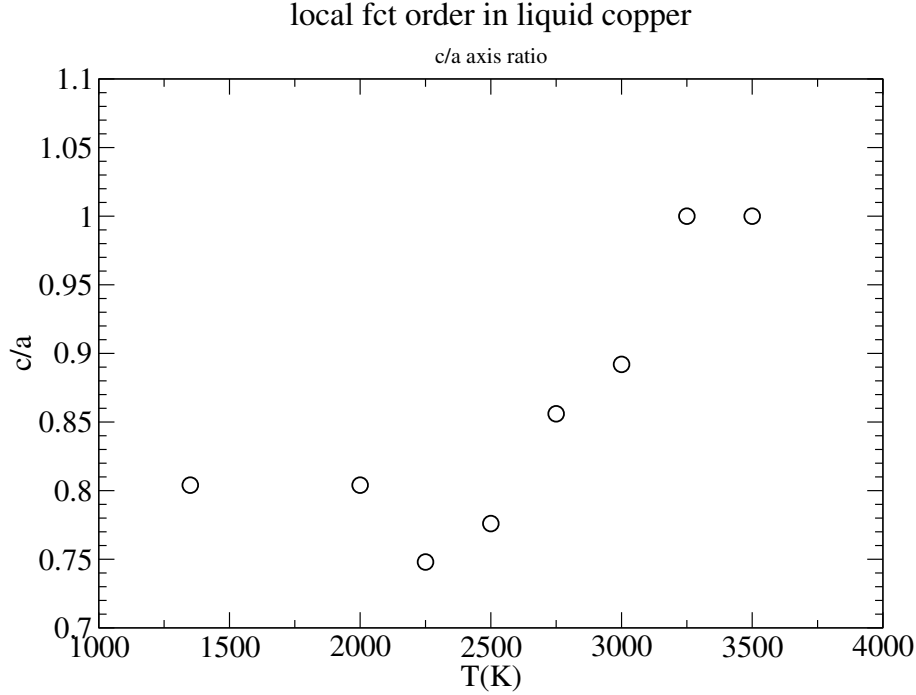}
    \caption{The figure presents the values of the axis length ratio $c/a$ giving the best agreement between our liquid-phonon calculations and experimental values.}
    \label{fig:Cu_resistivity2}
\end{figure}

Figure~\ref{fig:Cu_resistivity1} shows the electrical resistivity calculated for liquid copper, considering the persistence of a local order of the fct-type (blue triangles) and compares them to the experimental values \cite{Gathers1983}, represented by the black circles. Figure~\ref{fig:Cu_resistivity2} presents the values of the axis length ratio $c/a$ giving the best agreement between our liquid-phonon calculations and the experimental values. Our conclusions are similar to those we reached for liquid aluminum. With the exception of the values calculated at 2000 and 2500 K, we found that accounting for compression of the fcc cell along the vertical axis yields good agreement with experimental data. The values of the axis ratio $c/a$ are comparable to those found for liquid aluminum. The discrepancy observed at 2000 K and 2250 K could be due to errors in evaluating the mean ion charge at these temperatures.

\begin{table}[!ht]
    \centering
    \begin{tabular}{ccccccccc}
    \hline
     $\rho$ (g.cm$^{-3}$) & 7.832 & 7.516 & 7.319 & 7.086 & 6.921 & 6.764 & 6.565 & 6.423 \\
     T (K) & 1356 & 2000 & 2250 & 2500 & 2750 & 3000 & 3250 & 3500 \\
     $Z^*$ & 1.613 & 1.537 & 1.492 & 1.444 & 1.416 & 1.395 & 1.382 & 1.387\\
    \hline
    \end{tabular}
    \caption{Values of the mean ion charge $Z^*$ obtained for liquid copper with the average-atom code {\sc Paradisio} \cite{Wetta2025}.}
    \label{tab:Z*_Cu}
\end{table}

\subsection{Impact of the uncertainty in $\gamma_G$ on our results}

Proceeding on the same way as for the evaluation of the error on the isobaric heat capacities due to the uncertainty $\Delta\gamma_G/\gamma_G$, one obtains the following estimation for the error on the Debye-Waller factor:

\begin{equation*}
\dfrac{\Delta(2W)}{2W} \approx \dfrac{\Delta\gamma_G}{\gamma_G}\left\{ 6\dfrac{(\theta_F/\theta_D-1)}{(3-2\theta_F/\theta_D)} \ln\left(\dfrac{\theta_D}{\theta_D^0} \right) \right. \\
+\left.2\int_{T_m}^T \gamma_G\,\alpha_V\, dT\right\}.
\end{equation*}

$\Delta\gamma_G/\gamma_G$ is the highest close to melting. At this point of state, we found that $\Delta(2W)/2W\approx 0.25\,\Delta\gamma_G/\gamma_G$ for both liquid copper and liquid aluminum. For the latter: $\Delta(2W)/2W\approx 0.075$ for aluminum, and is negligible for copper. Assuming that only the shortest reciprocal vectors $G$ contribute to the additional contribution $\delta\rho_\mathrm{dc}$ to total resistivity in the liquid state, we estimate the impact of $\Delta(2W)/2W$ on the total resistivity $\rho_\mathrm{dc}=\rho_\mathrm{dc}^Z+\delta\rho_\mathrm{dc}$ as follows:
\begin{equation*}
\dfrac{\Delta\rho_\mathrm{dc}}{\rho_\mathrm{dc}}=\dfrac{\Delta\delta\rho_\mathrm{dc}}{\delta\rho_\mathrm{dc}}\dfrac{\delta\rho_\mathrm{dc}}{\rho_\mathrm{dc}^Z+\delta\rho_\mathrm{dc}},  
\end{equation*}
where:
\begin{equation*}
\dfrac{\Delta\delta\rho_\mathrm{dc}}{\delta\rho_\mathrm{dc}}\approx \dfrac{\Delta\mathrm{e}^{-2WG^2}}{\mathrm{e}^{-2WG^2}}\approx -2WG^2\,\dfrac{\Delta(2W)}{2W}. \end{equation*}
The error $\Delta\rho_\mathrm{dc}/\rho_\mathrm{dc}$ is greatest near the melting point and reduces as the liquid's temperature rises, due to the exponential decrease with $2WG^2$ of $\delta\rho_\mathrm{dc}$. Close to the melting temperature $2WG^2$ is less than unity and the error $\Delta\rho_\mathrm{dc}/\rho_\mathrm{dc}$ on the total resistivity reduces to a fraction of the one on the $\delta\rho_\mathrm{dc}$, \emph{i.e.}:
\begin{equation*}
\dfrac{\Delta\rho_\mathrm{dc}}{\rho_\mathrm{dc}}\lesssim -2WG^2\,\dfrac{\Delta(2W)}{2W}.    
\end{equation*}
For liquid aluminum near its melting point: $2WG^2\approx 0.4$, and accounting for $\Delta(2W)/2W\approx 0.075$, we estimate that the total resistivity error induced by uncertainty in the Gr\"uneisen parameter value falls within the 5\% experimental error announced by Gathers.

The question is whether adjusting the $r=c/a$ ratio can compensate for this error. The impact on the exponential $\mathrm{e}^{-2WG^2}$ is low, even negligible. A change $\Delta r$ of the axis ratio mostly affects the value of $\delta\rho_\mathrm{dc}$ through the lower bound of the integral in Eq.~(\ref{correction_resistivité}). Since the derivative of the Fermi-Dirac distribution function closely resembles a peak centered on the chemical potential $\mu$, even small changes in the lengths of the reciprocal vectors $G$ induced by small adjustments of the $c/a$ ratio have a noticeable effect on the integral's value. We estimate that the necessary adjustments $\Delta r/r$ to counterbalance the error in the resistivity resulting from uncertainty in the Gr\"uneisen parameter are of the order of a few percents.
 
\section{Conclusion}

In this study, we investigated the Debye-Waller factor in the liquid state using a formalism that combines liquid-phonon theory with key features of the Yukawa one-component model. Anharmonicity of the phonon vibrations was treated at the quasi-particle level. 

Liquid-phonon theory introduces the concept of phonon relaxation time, which is determined by the ratio of liquid shear viscosity to infinite-frequency shear modulus. Liquid phonon theory applies as long as $\theta_F<\theta_D$. Crossing the Frenkel line defined by $\theta_F=\theta_D$ leads to fundamental changes in pair correlations, both in reciprocal space and in real space, as well as in thermodynamics, scaling laws, and phonon states \cite{Bolmatov2015}. Coupling parameter  $\Gamma/\Gamma_m\approx 0.05$ marks the crossover. The liquid-phonon theory applies to values $0.05\lesssim \Gamma/\Gamma_m\leq 1$.

The main argument in favor of the YOCP model is that it belongs to the class of R-simple liquids. To qualify as R-simple, the correlation parameter $R$ of the thermal fluctuations $\Delta W-\Delta U$ must be at least 0.9. This condition is met for $1\leq \kappa \leq 5$ and $\Gamma>1$. For the most strongly coupled liquids ($\Gamma/\Gamma_m>0.1$), $R$ approaches 0.98.

As the temperature rises, it may be necessary to account for anharmonic effects beyond the quasi-particle approximation level. While the latter appears sufficient for liquid copper, it may not be for aluminum. This could explain the discrepancy between calculated and observed experimental $C_P(T)$ values for aluminum. However, this discrepancy could also be due to the difficulty of extracting a clear dependence of aluminum’s $C_P$ on $T$ from quasi-linear enthalpies.

All the quantities provided by the YOCP model, used in this work for the calculation of the ones needed by the liquid-phonon approach, are within the scope of current experimental techniques. Viscosity measurements are available for many liquid metals, including copper and aluminum \cite{Assael2006,Assael2010,Assael2012,Assael2018}. The high frequencies necessary for measuring the instantaneous shear modulus ($G_\infty$) are also accessible \cite{Dyre2012}. The ratio of $G_\infty$ on viscosity determines the lifetime of the phonon shear waves, directly accessible  through dynamic structure factors  obtained via X-ray scattering or neutron scattering experiments in the THz range \cite{Hosokawa2015}. Isomorphism theory relates the Gr\"uneisen parameter $\gamma_G$ to the exponent scale constant $\gamma$  determined by the quasi-proportionality between the isochoric fluctuations of the virial and potential energies exhibited by R-simple liquids \cite{Schroder2009}. The dynamic properties (\emph{e.g.}  relaxation time, diffusivity, or viscosity) of these systems are functions of $\rho^\gamma/T$, which provides a means of determining the exponent  and, consequently, of inferring relevant Gr\"uneisen parameters for dense liquids which account for the latter's elastic/viscous nature \cite{Gundermann2011}. 

We first applied liquid-phonon theory to the temperature dependence of the heat capacity at constant pressure in the liquid state. A comparison with the isobaric heat capacity reported by Gathers for liquid aluminum and liquid copper shows good agreement when persistence of local structural order in the liquid is assumed. We compared the quantities required by the theory and estimated using YOCP with the experimental data and found good agreement for the Frenkel temperature $\theta_F$ as well as for the thermal expansion parameter $\alpha_V$, both important in the formalism.  The only discrepancy concerns the Gr\"uneisen parameter in dense liquids. We analyzed the possible impact of this uncertainty on $C_P(T)$ and concluded that it is insufficient to call our results regarding the nature of the liquid into question. 
Encouraged by this result, we extended the formalism to the Debye-Waller factor in the liquid state. We arrived at an expression that explicitly introduces an additional temperature dependence resulting from the finite and temperature-dependent phonon lifetime in liquids.

The Debye-Waller factor is an essential quantity in our methodology for computing the electrical resistivity of dense matter based on the Ziman theory and average-atom model results. Our resistivity approach introduces a correction to the Ziman formula that accounts for the persistence of partial crystal-type ordering in these states of matter. The Debye-Waller factor determines the temperature dependence of this correction, ensuring continuity from the solid to the plasma state through the liquid state. We applied this methodology, with liquid-phonon Debye-Waller factors, to the interpretation of the electrical resistivities of liquid aluminum and liquid copper, as probed by Gathers.

We concluded that a local fct-type order persisted in both materials in the liquid state. The axis ratio was found to vary from $c/a \approx 0.8$ to $c/a \approx 1$ between melting and the highest temperature studied, with a fairly high degree of confidence in the copper results, and an error of approximately $\pm$ 0.06 for liquid aluminum, estimated on the basis of the discrepancy observed between the YOCP Gr\"uneisen  parameters and those derived from shock/release experiments.

Our findings suggest that resistivity measurements could be a valuable tool for investigating the nature and extent of crystalline order in liquids. This could provide new insights into the structural characterization of high-temperature liquid metals. Computer simulations have recently revealed that crystal-like pre-ordering plays a key role in the nucleation and growth mechanisms during crystallization. Understanding the physical mechanisms involved in crystallization is important in many fields, including metallurgy, drug production, nanoscale electronics and protein engineering. 

Over the past two decades, there has been a growing interest in complex fluids and plasmas \cite{Morfill2009}. This term encompasses a wide variety of systems, including glasses, melts, colloidal suspensions, dusty plasmas or granular medias. Most of them can be described through Yukawa potentials. The historical YOCP has been derived (among others) into 2D-YOCP, which finds  namely application in the study of dusty plasmas and colloidal suspensions \cite{Donko2008_complexplasmas}, or so-called ``Yukawa balls'' models for the description of spherical crystals in fluids \cite{Bonitz2010}. Describing and commenting on these new applications of the Yukawa potential is beyond the scope of the present work. However, one can guess that these approaches will benefit from the findings by the isomorph theory for the usual YOCP. This is already the case for the 2D-YOCP system, where isomorphism theory generalizes the Rosenfeld-Tarazona scaling law to dusty plasmas \cite{Castello2019}. 

%\bibliography{refs-Debye-Waller}

\end{document}